\documentclass[11pt,a4paper]{article}

\usepackage[utf8]{inputenc}
\usepackage[T1]{fontenc}
\usepackage[british]{babel}
\usepackage{amsmath,amssymb,bm}
\usepackage{empheq}
\usepackage{graphicx}
\usepackage{booktabs}
\usepackage{array}
\usepackage{siunitx}
\usepackage[dvipsnames]{xcolor}
\usepackage{tikz}
\usetikzlibrary{arrows.meta,positioning,calc,fit,backgrounds}
\usepackage[round,authoryear]{natbib}
\usepackage[margin=2.5cm]{geometry}
\usepackage{hyperref}
\hypersetup{colorlinks=true,linkcolor=RoyalBlue,citecolor=ForestGreen,urlcolor=BrickRed}

\newcommand{\Retau}{\mathit{Re}_{\tau}}
\newcommand{\Ubulk}{U_{b}}
\newcommand{\dpdx}{\partial_{x} p}
\newcommand{\ww}{w_{w}}
\newcommand{\tauw}{\tau_{w}}
\newcommand{\Epump}{\mathcal{P}_{p}}
\newcommand{\Ww}{\mathcal{W}_{w}}
\newcommand{\Win}{\mathcal{W}_{\mathrm{in}}}
\newcommand{\Eps}{\varepsilon}
\newcommand{\DR}{\mathrm{DR}}
\newcommand{\avg}[1]{\left\langle #1 \right\rangle}

\definecolor{myblue}{RGB}{31,119,180}
\definecolor{myorange}{RGB}{255,140,0}
\definecolor{ggreen}{RGB}{30,140,60}
\definecolor{rred}{RGB}{200,40,40}
\definecolor{flowblue}{RGB}{0,102,204}
\definecolor{transblue}{RGB}{100,150,255}

\title{\vspace{-1.5em}Reward hacking in physical reinforcement learning revealed by turbulent drag reduction}
\author{
G.~M. Cavallazzi$^{1,*}$, M. P\'erez Cuadrado$^{1}$, A. Pinelli$^{1}$
\\[4pt]
\small $^{1}$School of Science and Technology, Department of Engineering,\\
\small City St.\ George's, University of London, London, UK\\
\small $^{*}$Corresponding author: \texttt{giorgio.cavallazzi@city.ac.uk}
}
\date{}

\begin{document}
\maketitle

% =======================================================================
\begin{abstract}
\noindent
Reinforcement-learning controllers optimise specified rewards, but in physical systems those rewards often capture only part of the true control objective. Three mechanisms through which this mismatch can produce apparent success without physical improvement are identified: incomplete accounting that omits relevant costs, constraint enforcement outside the policy that corrupts credit assignment, and observations that fail to resolve the relevant dynamics. All three are demonstrated in active drag reduction of wall-bounded turbulence, where the conservation constraint and full energy budget can be measured directly. A memoryless learnt policy reports drag reduction while raising total dissipation, collapsing to non-physical flow configurations. A recurrent multi-agent controller with the zero-mean projection embedded in the actor, temporal memory matched to the relevant timescales, and an actuation cost that bounds the wall power delivers a physically consistent control. Progress in physical reinforcement learning requires the reward, constraints, observations and evaluation metrics to represent unequivocally the physical objective.
\end{abstract}

%%%%%%%%%%. INTRO %%%%%%%%%%%%%%
Reinforcement-learning agents are trained to maximise a specified reward signal \citep{sutton2018rl}. In many physical control problems, however, that reward is only an imperfect surrogate for the target quantity. When the reward omits important aspects of the desired outcome, optimisation can favour behaviours that improve the training objective without improving the underlying physical objective. Related concerns have been discussed extensively in the machine-learning literature under labels such as reward hacking, specification gaming and reward misspecification \citep{amodei2016aisafety,skalse2022hacking}. A proxy is hackable in the formal sense of \citet{skalse2022hacking} when two policies exist that the proxy and the true objective rank in opposite orders, which is exactly the situation documented below.

As reinforcement learning has moved from games and simulated benchmarks to the control of physical systems \citep{rabault2019cylinder,garnier2021reviewdrl,vignon2023review}, these concerns have acquired additional significance because rewards, constraints and observations are no longer independent but are linked through the conservation laws and dynamical evolution of the process being controlled.

Conservation laws can alter the relationship between actions and rewards, constraints may be enforced outside the optimisation loop, and observations may only partially resolve the state of the system. At the same time, the quantities used to evaluate performance are often proxies for the true physical objective and may omit important energetic costs. As a result, improvements in the reported reward do not necessarily imply improvements in the target quantity.

In particular, two features of physical control sharpen the difficulty. The actuation is
rarely free to take any value: conservation laws of mass or of charge tie
the admissible actions together, so that what one part of a distributed
actuator may do depends on what the rest of it is doing. When the control
is shared among many local agents, a scalable and now common
arrangement, this coupling cuts against the separable per-agent
contributions that the learning signal is built to credit
\citep{foerster2018coma}. Physical systems also carry their own time
scales. A policy that maps an instantaneous measurement to an action
follows a fast process but loses the phase of a slow one
\citep{kaelbling1998pomdp,hausknecht2015drqn}, and the quantity worth
controlling is often the slow one.

Active control of wall-bounded turbulence provides a useful setting in which to examine these issues. The objective is well defined: reducing the total energetic cost associated with skin-friction drag, which accounts for much of the energy spent moving vehicles and pumping fluids
through pipes. Cutting it even by a few percent carries a large
economic and environmental payoff \citep{spalart2011dragreduction}. In this framework, the governing equations are known, the conservation constraints are explicit, and the energy budget can be measured directly. 
At the same time, the flow contains a hierarchy of interacting spatial and temporal scales that make it a challenging control problem. These scales emerge from the near-wall regeneration cycle that sustains turbulence production and ultimately determines the skin friction a controller seeks to reduce.
The cycle consists of the continual regeneration of streaks and quasi-streamwise vortices \citep{kline1967structure,hamilton1995regeneration,waleffe1997ssp} and operates autonomously within roughly one hundred wall units of the surface \citep{jimenez1999autonomous}.
Lengths here are quoted in wall units, physical
distances non-dimensionalised by the near-wall viscous scale
$\nu/u_\tau$ formed from the skin-friction velocity $u_\tau$, so that the
channel height in wall units is the friction Reynolds number $\Retau$
built on the wall shear stress. Model-based wall control has a long
record against
this cycle, from opposition control \citep{choi1994opposition}, which
removes roughly a quarter of the drag at comparable Reynolds number,
through suboptimal and predictive laws returning $16$ to $22\%$
\citep{lee1998suboptimal,bewley2001predictive}. Open-loop spanwise
forcing reaches further on drag, and the literature that developed it has
reported the net budget rather than the drag alone throughout:
streamwise-travelling waves of spanwise velocity save at most $18\%$ of
the pumping power once the actuation is charged, steady waves $5\%$, and
for several open-loop techniques the net saving is negative
\citep{quadrio2009stw}. The net energy budget rather than the nominal
drag is thus the established basis for ranking
schemes once the actuation cost is counted
\citep{fukagata2009netpower,marusic2021energypath}. Reinforcement
learning re-entered through the channel in a multi-agent form, an
identical policy on every wall patch with a centralised critic during
training (centralised training, decentralised execution, CTDE)
\citep{vignon2023rb,guastoni2023drl}, and the template that has spread
since shares a parameter-shared CTDE policy, the drag-reduction
percentage as the objective, and the reported figure as the headline
\citep{sonoda2023rl,cavallazzi2024wallcycle,walchli2024minimalchannel}.

That template inherits three problems that the headline figure does not
expose. The first is structural to the multi-agent formulation. Blowing
and suction must inject no net mass: a net flux through the wall is
incompatible with incompressibility in the closed channel, and even held
numerically stable it would add momentum to the flow, lowering the drag
by simply moving around terms in the momentum balance. The joint action is
therefore projected onto its zero-mean subspace before it reaches the
flow; under that projection an agent's applied action depends on what
every other agent did, and the reward an agent should be credited for
can no longer be separated from its neighbours'. The second is a question of
observability. The buffer-layer cycle evolves over roughly a hundred
viscous time units, while the policy acts on an instantaneous slice; a
memoryless map from a single snapshot cannot represent the phase of a
cycle that is slow compared with its own sampling. The third concerns
the objective. In constant-flow-rate operation the drag-reduction
percentage measures only the saved pumping power and ignores the power
the actuation delivers to the fluid. The standard accounting in the
learning literature charges that actuation a kinetic-energy-flux cost
that scales with the cube of the actuation amplitude, which is then
found negligible \citep{kametani2015blowingsuctionTBL,guastoni2023drl};
the thermodynamic work the wall actually does on the flow is a different
quantity, and one that the classical control literature has long
insisted on when ranking schemes by net energy budget
\citep{fukagata2009netpower,marusic2021energypath}.

\begin{figure}[htbp]
\centering
\resizebox{\linewidth}{!}{%
\begin{tikzpicture}[every node/.style={font=\footnotesize}, >=Stealth,
   axis/.style={thick,->},
   badge/.style={circle,draw,fill=ggreen!35,inner sep=0.8pt,font=\scriptsize\bfseries},
   fix/.style={draw,rounded corners,fill=ggreen!10,align=center,inner sep=4pt,font=\scriptsize,text width=42mm}]
  % --- physical channel: wall with blowing/suction and the field above it ---
  \begin{scope}[xshift=-5.5cm, x={(1,-.25,-.25)}, y={(0,1,0)}, z={(.25,0,-1.25)}, scale=0.42]
    \input{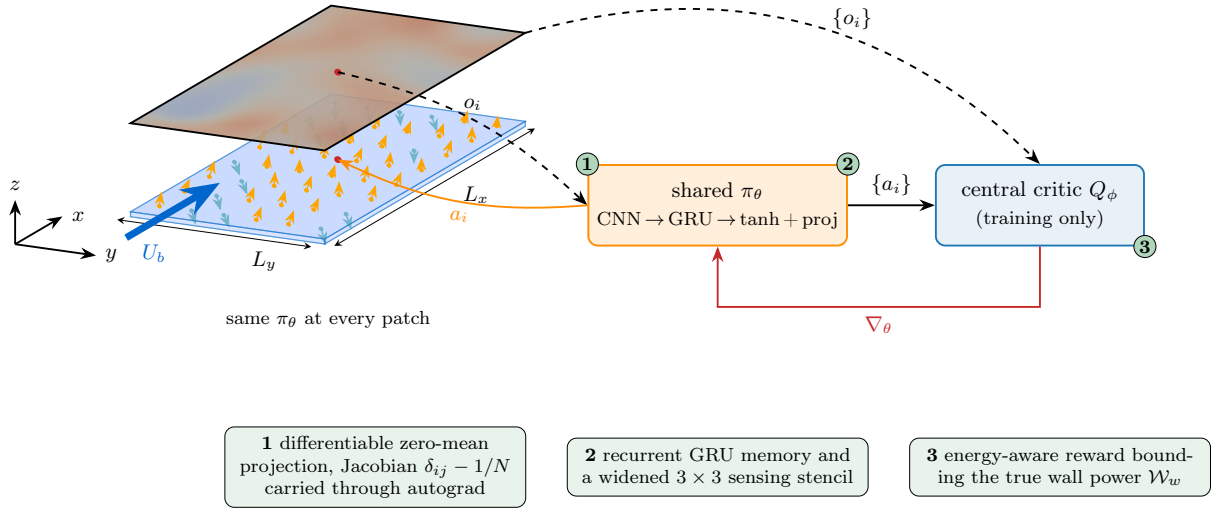}
  \end{scope}
  \node at (-4.0,-1.4) {\scriptsize same $\pi_\theta$ at every patch};
  % --- shared policy and central critic ---
  \node[draw=myorange, fill=myorange!15, thick, rounded corners,
        minimum width=3.1cm, minimum height=1.2cm, align=center] (pi) at (1.8,0.3)
        {shared $\pi_\theta$\\[1pt]\scriptsize CNN\,$\to$\,GRU\,$\to$\,$\tanh$\,+\,proj};
  \node[draw=myblue, fill=myblue!10, thick, rounded corners,
        minimum width=3.1cm, minimum height=1.2cm, align=center] (Q) at (6.6,0.3)
        {central critic $Q_\phi$\\[1pt]\scriptsize (training only)};
  % --- closed loop: observe, act, critique, learn ---
  \draw[->, thick, dashed]   (UPPER_P)  to[bend left=15] node[pos=0.5,above]{$o_i$} (pi.west);
  \draw[->, thick, myorange] (pi.west)  to[bend left=15] node[pos=0.5,below]{$a_i$} (WALL_P);
  \draw[->, thick]           (pi.east)  -- node[above]{$\{a_i\}$} (Q.west);
  \draw[->, thick, dashed]   (UPPER_TR) to[bend left=30] node[pos=0.6,above]{$\{o_i\}$} (Q.north);
  \draw[->, thick, rred]     (Q.south)  -- ++(0,-0.9) -| node[below,pos=0.25,rred]{$\nabla_\theta$} (pi.south);
  % --- numbered safeguards placed on the loop ---
  \node[badge] at (pi.north west) {1};
  \node[badge] at (pi.north east) {2};
  \node[badge] at (Q.south east)  {3};
  % --- the three fixes spelled out below ---
  \node[fix] (f1) at (-3.3,-3.6) {\textbf{1} differentiable zero-mean projection, Jacobian $\delta_{ij}-1/N$ carried through autograd};
  \node[fix] (f2) at (1.8,-3.6)  {\textbf{2} recurrent GRU memory and a widened $3\times3$ sensing stencil};
  \node[fix] (f3) at (6.9,-3.6)  {\textbf{3} energy-aware reward bounding the true wall power $\Ww$};
\end{tikzpicture}}
\caption{Setup and the corrected control loop. A turbulent channel at
$\Retau\simeq180$ is driven at constant flow rate, and the lower wall is
actuated by zero-net-mass blowing and suction (orange out, teal in) on a
grid of patches; the field above the wall is the streamwise velocity on
the detection plane the policy reads. Every patch runs the same actor
$\pi_\theta$, a convolutional encoder feeding a GRU and a $\tanh$ head
with the zero-mean projection as its last layer, reading a $3\times3$
ring of patches $o_i$ and emitting the wall-normal action $a_i$; a
central critic $Q_\phi$ sees the joint state and action during training
only and returns the policy gradient $\nabla_\theta$. The three numbered safeguards correct, in turn, the corrupted credit
assignment caused by enforcing mass conservation outside the actor, the
partial observability of the slow near-wall cycle, and the reward
incentive to reduce drag by pumping power through the wall.}
\label{fig:overview}
\end{figure}

We treat these as design faults with concrete fixes and report the
controller that results (Fig.~\ref{fig:overview}). A differentiable
projection layer puts the zero-mean constraint inside the actor, so the
policy gradient sees the constraint rather than fighting it after the
fact. A recurrent core and a widened sensing stencil give the policy the
memory the cycle demands. An energy-aware reward, whose actuation cost
bounds the physically correct wall power, removes the incentive to pump
the wall.
Two deliberately degenerate controllers, an open-loop stripe pattern
and a memoryless learnt policy with none of the fixes, show what the
unguarded objective rewards: both post sizeable drag-reduction
percentages while raising the total dissipation above the uncontrolled
value, and the memoryless learnt policy reaches that state by collapsing
onto a wave at the box wavelength, a reward-hacking artefact rather
than control of the flow.
The corrected controller, a recurrent multi-agent policy we refer to
throughout as GRU-MARL, instead reduces the drag it is supposed to, at
an energy budget that matches opposition and at an amplitude well below
it, and it transfers from its small training domain to a much larger
evaluation channel without retraining.

% =======================================================================
\section*{Results}

\subsection*{Reported drag reduction can conflict with the physical objective}

All controllers are evaluated in the same constant-flow-rate
half-channel at $\Retau\simeq180$, on a box of
$(L_x^+,L_y^+,H^+)\simeq(1922,576,180)$ wall units actuated through the
wall-normal velocity alone, with the zero-net-mass constraint
\begin{equation}
  \int_{\Gamma_w}\ww(x,y,t)\,\mathrm{d}x\,\mathrm{d}y = 0
  \qquad\forall\,t
  \label{eq:zeromass}
\end{equation}
enforced at every step. In this setting the bulk kinetic-energy budget
closes on two external power terms, the pumping power supplied by the
flow-rate controller, per unit volume, and the work the actuation does on
the fluid, per unit wall area,
\begin{equation}
  \Epump = \lvert\avg{\dpdx}\rvert\,\Ubulk,
  \qquad
  \Ww = -\frac{1}{L_x L_y}\int_{\Gamma_w}\avg{\ww\,p}\,\mathrm{d}x\,\mathrm{d}y .
  \label{eq:powers}
\end{equation}
The flow rate is held constant by a uniform body force whose magnitude
equals the instantaneous mean streamwise pressure gradient, so the
pressure field solved for carries no mean-gradient contribution and its
volume mean is set to zero, the equations of motion depending on the
pressure only through its gradient. $\Ww$ is in any case independent of
that choice of datum, since~\eqref{eq:zeromass} makes the wall-integrated
actuation vanish at every step. The invariance holds to numerical
precision: across the controllers the wall-pressure datum differs by
several units, yet removing it changes the measured $\Ww$ by at most one
part in $10^{6}$, and the discrete zero-net-mass residual stays below
$10^{-8}$ of the mean actuation magnitude.
The viscous wall-traction work vanishes identically here: continuity at
a wall that is no-slip in the wall-parallel components forces
$\partial_z w|_{\mathrm{wall}}=0$ pointwise, so the only surviving
wall-work term is the pressure covariance in~\eqref{eq:powers}.
Stationarity then fixes the volume-averaged dissipation,
\begin{equation}
  \Eps = \Epump + \frac{\Ww}{H},
  \label{eq:balance}
\end{equation}
with the factor $1/H$ converting surface work to a volume-distributed
rate. The quantity a useful controller must lower is $\Eps$. The
conventional benchmark instead reports
\begin{equation}
  \DR = 1 - \frac{\lvert\avg{\dpdx}\rvert}{\lvert\avg{\dpdx}_0\rvert},
  \label{eq:dr}
\end{equation}
the relative saving in pumping power, which says nothing about $\Ww$.
Subtracting~\eqref{eq:balance} from its uncontrolled value separates the
two contributions,
\begin{equation}
  \Delta\Eps = \Eps_0 - \Eps
  = \bigl(\lvert\avg{\dpdx}_0\rvert\,\Ubulk - \Epump\bigr) - \frac{\Ww}{H}
  = \DR\,\lvert\avg{\dpdx}_0\rvert\,\Ubulk - \frac{\Ww}{H},
  \label{eq:netgain}
\end{equation}
so a controller can raise $\DR$ by pumping the wall rather than by
quieting the flow. In the last form the loophole is explicit:
$\Delta\Eps$ and $\DR$ agree only while $\Ww$ stays negligible.

This loophole is usually argued away with the wrong cost. The standard
accounting in the blowing/suction learning literature charges the
actuation a kinetic-energy-flux input power
\begin{equation}
  \Win = \frac{1}{2}\,\frac{1}{L_x L_y}\int_{\Gamma_w}\avg{\lvert\ww\rvert^{3}}\,\mathrm{d}x\,\mathrm{d}y,
  \qquad
  S = \frac{c_{f,0} - (c_f + \Win)}{c_{f,0}},
  \label{eq:proxy}
\end{equation}
and reports the net-energy saving $S$ alongside the drag reduction
\citep{kametani2015blowingsuctionTBL,guastoni2023drl}.
Because $\Win$ scales as the cube of the actuation amplitude, at the
$\mathcal{O}(10^{-2})$ r.m.s.\ amplitudes of the smooth controllers it
evaluates to $\mathcal{O}(10^{-6})$, three orders below the friction
coefficient, and the net saving collapses onto the drag reduction. For a
saturated policy it is larger, but it remains a function of the actuation
amplitude alone,
so penalising it merely caps the output; it cannot register the
covariance $\avg{\ww\,p}$ through which a bounded-amplitude actuation
still pumps real power into the flow. The true wall power
in~\eqref{eq:powers} is that covariance, and it is what enters the
dissipation balance.

Table~\ref{tab:energy} measures every term of~\eqref{eq:balance}
directly for five controllers spanning the design space, and the
amplitude proxy and the true wall power part company exactly where it
matters. The proxy is not blind to the saturated policies: it charges the
stripes and the memoryless vanilla-DRL policy $1.05\times10^{-3}$ and
$6.33\times10^{-4}$, some thirteen and eight percent of the uncontrolled
friction coefficient. It is nonetheless too small, by factors of $1.8$
and $4.6$ against the true wall power, and that shortfall is what decides
the sign: carried through~\eqref{eq:proxy} the proxy still reports net
savings of $+20\%$ and $+8\%$ for these two. The peak amplitude
$\lvert\ww\rvert_{\max}$ is no better a guide, being essentially the
same, between $0.118$ and $0.131$, across opposition, the stripes and
vanilla DRL, while their true wall power
$\Ww$ spans $7\times10^{-6}$ to $2.91\times10^{-3}$, a factor of roughly
four hundred. The open-loop
stripe pattern records the highest drag-reduction percentage in the
table, $33.2\%$, while driving $\Eps$ fourteen percent above the
uncontrolled value; the memoryless vanilla-DRL policy reports $15.5\%$
drag reduction while its wall-work lifts the total dissipation by more
than half. Both are nominal successes and physical failures, and
neither $\DR$ nor the amplitude proxy can separate them from a genuine
reduction. Opposition and GRU-MARL sit on the other side
of~\eqref{eq:netgain}: their wall-work is two to three orders of
magnitude smaller, so the wall is a near-passive observer in the sense
of net energy injection and their ordering by $\Eps$ follows their
ordering by $\DR$.

\begin{table}[h]
\centering\small
\renewcommand{\arraystretch}{1.25}
\setlength{\tabcolsep}{6pt}
\begin{tabular}{@{}lrrrrrrr@{}}
\toprule
Controller & $\DR$\,[\%] & $\Epump\,(\times10^{3})$ &
$\Win\,(\times10^{3})$ & $\Ww\,(\times10^{3})$ & $\Eps\,(\times10^{3})$ &
$\Delta\Eps\,[\%]$ & $\lvert\ww\rvert_{\max}$ \\
\midrule
uncontrolled        & $0.0$  & $4.10$ & $0.0000$ & $0.00$ & $4.10$ & $0.0$    & $0.000$ \\
opposition          & $21.4$ & $3.22$ & $0.0008$ & $0.007$ & $3.22$ & $+21.5$  & $0.131$ \\
stripes (open-loop) & $33.2$ & $2.74$ & $1.05$ & $1.93$  & $\mathbf{4.67}$ & $\mathbf{-13.9}$ & $0.128$ \\
vanilla DRL         & $15.5$ & $3.46$ & $0.633$ & $2.91$  & $\mathbf{6.38}$ & $\mathbf{-55.5}$ & $0.118$ \\
GRU-MARL            & $17.3$ & $3.39$ & $0.0016$ & $0.007$ & $3.39$ & $+17.3$  & $\mathbf{0.052}$ \\
\bottomrule
\end{tabular}
\caption{Energy budget of the five controllers, half-channel
constant-flow-rate units with $\Ubulk=H=1$. $\Delta\Eps$ is
$(\Eps_0-\Eps)/\Eps_0\times100$; positive marks a fall in total
dissipation, negative a net rise. The peak amplitude
$\lvert\ww\rvert_{\max}$ is nearly equal across opposition, stripes and
vanilla DRL, while the wall power $\Ww$ separates them by a factor of
$\sim\!400$. The amplitude proxy $\Win$ of~\eqref{eq:proxy} falls below
$\Ww$ in every case, by factors of $1.8$ and $4.6$ for the two saturated
policies, which is enough to reverse the sign of the net-energy verdict.
Bold marks the two controllers that raise $\Eps$ while reporting
positive $\DR$, and the smallest peak amplitude. Entries are quoted to
three significant figures, except $\Ww$ for opposition and GRU-MARL,
which is given to one; the two are not equal, differing beyond the
displayed precision.}
\label{tab:energy}
\end{table}

The stripe pattern earns its place in the table as a control: it is the
simplest policy that games the metric. It imposes a fixed square wave of
blowing and suction that alternates along the streamwise direction and is
uniform across the span,
\begin{equation}
  \ww(x,y) = w^{\star}\,\mathrm{sgn}\!\left[\sin\!\left(\frac{\pi x}{\Delta_s}\right)\right],
  \qquad \text{independent of } y \text{ and } t,
  \label{eq:stripes}
\end{equation}
with stripe half-period $\Delta_s$ at the smallest admissible patch size
and amplitude fixed at the saturation level $w^{\star}$. There is no
sensor and no feedback. It is a deliberately unintelligent open-loop
forcing, and it reaches the highest nominal drag reduction in the table
by injecting a large, steady wall power. The reason it belongs next to
the learnt controllers is empirical, and is the subject of the next
sections: the memoryless vanilla-DRL policy converges to behaviour of
the same kind.

A note on the evaluation domain. The drag-reduction and energy figures
reported here are all measured on the large box. A trained policy
evaluated in a minimal flow unit can drive the controlled flow towards
relaminarisation, at which point the pressure gradient drops to its
laminar value and the drag-reduction percentage saturates at an
artificially large number that reflects the box rather than the
controller. The large box keeps the flow turbulent throughout, so the
numbers in Table~\ref{tab:energy} are representative and are compared
across controllers at identical domain size and resolution.

\subsection*{Enforcing constraints outside the policy corrupts credit assignment}

\begin{figure}[htbp]
\centering
\begin{tikzpicture}[font=\footnotesize,>=Stealth,
   box/.style={draw,rounded corners,minimum height=9mm,minimum width=20mm,align=center}]
  % ----- panel a (top): projection after the actor -----
  \node[align=right,text width=24mm,font=\footnotesize] (la) at (-2.6,1.0)
       {\textbf{a}\\projection after\\the actor};
  \node[box,fill=myorange!15] (pa)    at (1.2,1.0) {$\pi_\theta\!\to\! a_i$};
  \node[box,fill=gray!12]     (proja) at (4.8,1.0) {$a'_i=a_i-\bar a$};
  \node[box,fill=myblue!12]   (enva)  at (8.4,1.0) {flow, reward};
  \draw[->] (pa) -- (proja);
  \draw[->] (proja) -- (enva);
  % gradient routed ABOVE the blocks
  \draw[->,rred,thick] (enva.north) |- ++(0,0.85)
        -| node[pos=0.25,above]{$\nabla$ on $a_i$, not $a'_i$} (pa.north);
  % ----- panel b (bottom): projection inside the actor -----
  \node[align=right,text width=24mm,font=\footnotesize] (lb) at (-2.6,-2.0)
       {\textbf{b}\\projection inside\\the actor};
  \node[box,fill=myorange!15] (pb)    at (1.2,-2.0) {$\pi_\theta$};
  \node[box,fill=ggreen!15]   (projb) at (4.8,-2.0) {$a'_i=a_i-\bar a$\\(last layer)};
  \node[box,fill=myblue!12]   (envb)  at (8.4,-2.0) {flow, reward};
  \draw[->] (pb) -- (projb);
  \draw[->] (projb) -- (envb);
  \draw[->,ggreen,thick] (envb.north) |- ++(0,0.85)
        -| node[pos=0.25,above]{$\nabla$ carries $\delta_{ij}-1/N$} (pb.north);
\end{tikzpicture}
\caption{Credit assignment under the zero-mean constraint. (a) With the
projection~\eqref{eq:proj} after the actor, the gradient is computed for
the emitted $a_i$ while the flow responded to the mixed $a'_i$, so
neighbouring agents contaminate each agent's credit. (b) Made the
actor's last layer, the projection's Jacobian~\eqref{eq:jac} propagates
during training and the gradient is taken on the applied field.}
\label{fig:credit}
\end{figure}
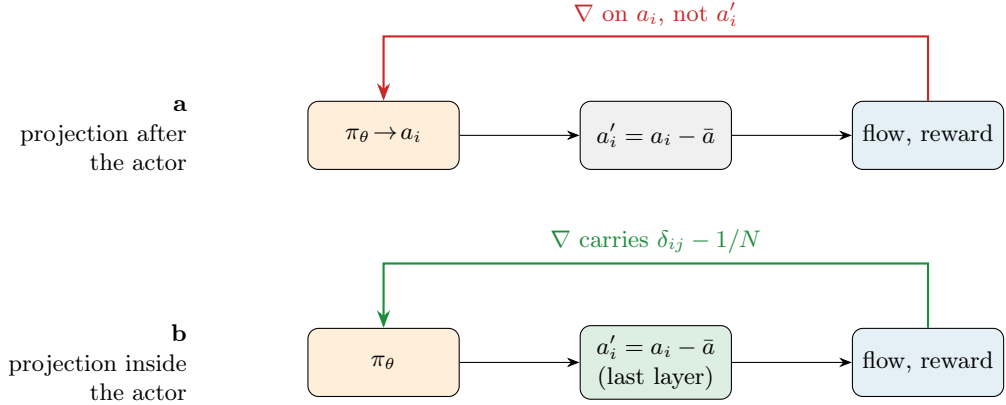

The wall is tiled into patches, each running an identical policy
$\pi_\theta$ in the parameter-shared multi-agent arrangement, with a
single centralised critic during training. Zero net mass forces the
scalar outputs $a_i$ onto their mean,
\begin{equation}
  a'_i = a_i - \frac{1}{N}\sum_{j} a_j,
  \label{eq:proj}
\end{equation}
before they are lifted to the wall. The flow receives $a'_i$, not $a_i$,
so the action that earns reward at patch $i$ is a function of every
other patch's output: a small $a_i$ can be turned negative by the mean
of its neighbours, and an action uniform across patches is sent to zero.
Applied as a post-processing step on the emitted actions, the projection
leaves the actor receiving a gradient computed for $a_i$ while the
environment responded to $a'_i$, and the per-agent credit the
deterministic policy gradient relies on is corrupted by the very
constraint that makes the actuation admissible
(Fig.~\ref{fig:credit}a).

The fix is to make~\eqref{eq:proj} the last layer of the actor. The
projection is linear, with constant Jacobian
\begin{equation}
  \frac{\partial a'_i}{\partial a_j} = \delta_{ij} - \frac{1}{N},
  \label{eq:jac}
\end{equation}
so automatic differentiation propagates the coupling back through the
network at no modelling cost. The actor learns to emit actions already
close to zero-mean, and the policy gradient is taken with respect to the
field the flow actually sees (Fig.~\ref{fig:credit}b). The mechanism is
not specific to the flow. In a fluid-dynamics-free model, with no partial
differential equation, no wall and no actuation energetics, a shared policy
controlling a set of agents under the same zero-mean projection learns
down to the reachable optimum when the Jacobian~\eqref{eq:jac} is
differentiated through, but stalls with most of its agents pinned against
their output bound when the projection is applied after the actor, the
same saturation the wall policy shows. The shortfall grows with the net
demand the conserved resource cannot supply and closes when that demand
is zero, which fixes the constraint itself, not any dynamics, as the
cause (Supplementary Fig.~S2).

\subsection*{Partial observability drives saturated actuations}

\begin{figure}[htbp]
\centering
\begin{tabular}{@{}c@{\hspace{6pt}}c@{}}
\includegraphics[width=0.46\textwidth,height=0.36\textheight,keepaspectratio]{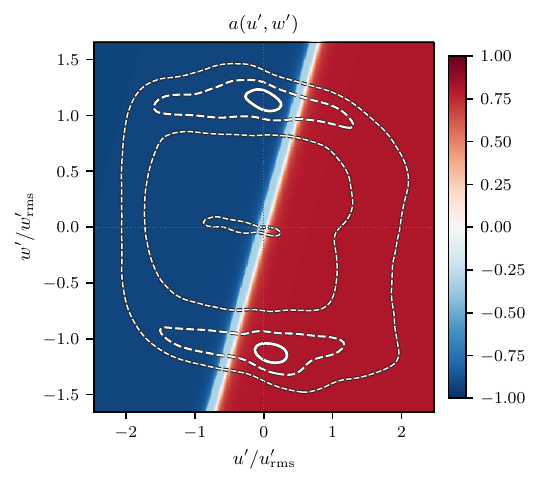} &
\includegraphics[width=0.52\textwidth,height=0.36\textheight,keepaspectratio]{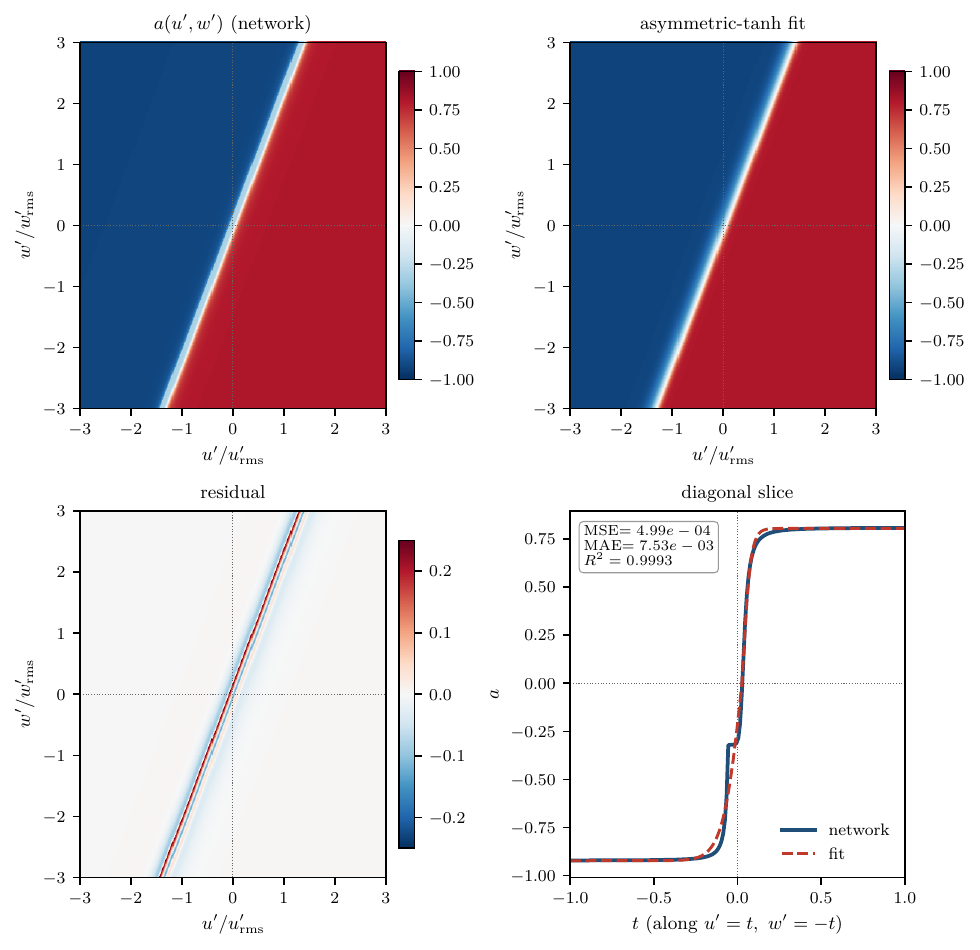} \\
{\footnotesize \textbf{a} vanilla DRL: $a(u',w')$} &
{\footnotesize \textbf{b} asymmetric 1-D $\tanh$ fit, $R^2\!\simeq\!0.999$} \\
\end{tabular}
\caption{The memoryless vanilla-DRL policy saturates. (a) Its action
surface over the two sensed scalars at the detection plane, overlaid with
the contours of the observed $(u',w')$ density (the white lines enclose
the bulk of the sampled states), so that the saturated region the policy
actually visits is visible. (b) The surface is reproduced by a saturating
$\tanh$ on a single linear combination $a_1 u'+a_2 w'$, so the policy is
effectively a one-dimensional switch pinned against its amplitude bound.}
\label{fig:saturation}
\end{figure}

Reinforcement learning casts control as a Markov decision process: the
action is chosen from the current state on the assumption that the state
and the action together determine what comes next. A memoryless policy
takes this literally, treating the instantaneous sensed slice as the full
state. The near-wall flow does not comply. 
The buffer-layer cycle turns over on a scale of order one hundred viscous time units, whereas a memoryless policy acts on an instantaneous slice of the flow; a single plane is therefore only a partial observation, and the current slice is not a sufficient statistic for the cycle the controller acts on.
%The buffer-layer cycle turns
%over on a scale of order one hundred viscous time units, far longer than the
%snapshot the policy reads at each step, so a single plane is a partial
%observation and the current slice is not a sufficient statistic for the
%cycle the controller acts on. 
A fixed map from that slice cannot recover
the phase of a process slow relative to its sampling, and the optimum it
settles on is degenerate. What that looks like is visible in the
vanilla-DRL actor, whose input is the pointwise pair of streamwise and
wall-normal fluctuations $(u',w')$ at the detection plane and whose
output is a scalar action $a=\tanh f_\theta(u',w')$. Its entire policy is
a surface on that plane, and the surface collapses onto an asymmetric
one-dimensional switch: a saturating $\tanh$ on a single linear
combination $a_1 u' + a_2 w'$ reproduces the trained network to
$R^2\simeq0.999$ (Fig.~\ref{fig:saturation}). The policy has degenerated
into a hard switch pinned against its amplitude bound, the saturated
two-level regime that a continuous-control return is known to favour when
nothing penalises it \citep{seyde2021bangbang}, and the behaviour the
wall-work column of Table~\ref{tab:energy} charges for. Saturated
switching is not by itself a defect, and for a range of control problems
it is optimal \citep{seyde2021bangbang}; what disqualifies this policy is
the state it drives the flow into, and the scales that state selects.

That the collapse follows from the timescale gap, and not from anything
in the flow, can be shown in a system with separated scales but no
turbulence. In a two-scale Lorenz--96 model, a control forced under the
same zero-mean projection and rewarded for lowering the slow-variable
energy reproduces both degeneracies: an actuation interval at the fast
decorrelation time leaves successive observations uncorrelated and drives
the policy to the same saturated two-level switching, while an interval
longer than the slow turnover acts on stale information and settles to
near-zero output. Only an intermediate cadence, short against the slow
scale yet long against the fast one, yields a structured low-amplitude
control, the surrogate of the interval $\Delta t_a^+\simeq5$ chosen at
the wall (Supplementary Fig.~S3).

\begin{figure}[htbp]
\centering
\includegraphics[width=\linewidth,height=0.82\textheight,keepaspectratio]{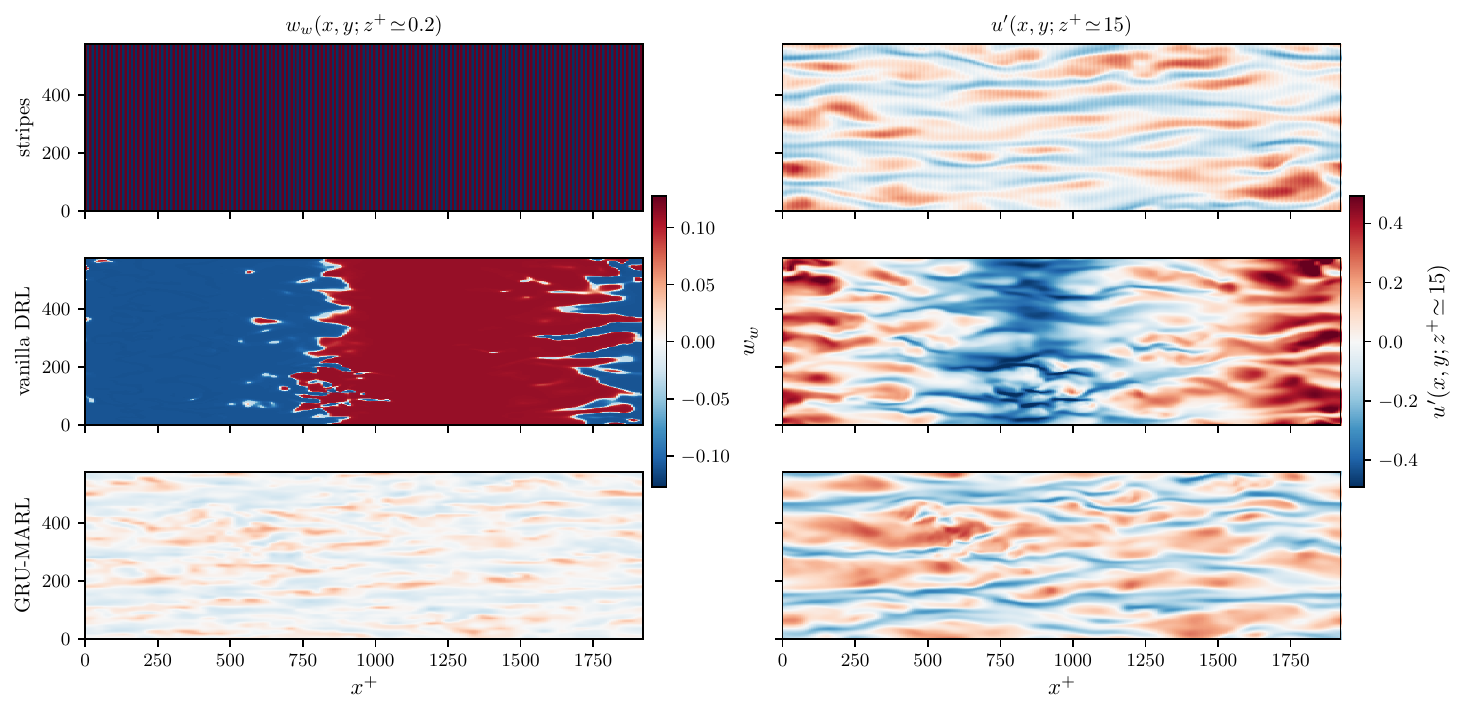}
\caption{Instantaneous wall actuation $\ww(x,y)$ (left column of each
row) and detection-plane streamwise fluctuation $u'$ (right column) on
the large box, in viscous units, for stripes, vanilla DRL and GRU-MARL.
Stripes is the imposed open-loop forcing, fixed by construction. Vanilla
DRL, though closed-loop, has settled into a wave of its own at the
longest wavelength the box admits, a learnt reward-hacking artefact
rather than control of the
flow, while GRU-MARL places its actuation in register with the
instantaneous streaks. Shared colour scale per column, uncapped in every
panel; the stripe row differs in appearance from the others because the
open-loop forcing is a two-level square wave that occupies the extremes
of that scale, not because it is rendered differently.}
\label{fig:snapshot}
\end{figure}

In the channel this saturated forcing takes a particular spatial form.
Figure~\ref{fig:snapshot} shows the instantaneous wall actuation for the
three actuated controllers on the large box; vanilla DRL, although a
closed-loop policy that reads the flow at every step, settles into a
banded field with no counterpart in the flow it is meant to control. Its
two-dimensional actuation spectrum carries $79\%$ of the wall-velocity
energy in a single mode, $k_x=1$ and $k_y=0$, uniform in span and at the
longest streamwise wavelength the periodic domain admits,
$\lambda_x=L_x$, whose phase advances by roughly $130^{\circ}$ per
actuation step. Both scales of the converged forcing are therefore
selected by the computation, the length by the box and the period by the
sampling interval, and neither by the near-wall cycle. We read the
particular mode as a problem-specific outcome rather than a universal
signature of the timescale failure. The high-frequency switching injects
an outsized amount of power into the domain, the wall-work column of
Table~\ref{tab:energy} measures it, and the box-scale wave is
the flow state this solver settles into under a forcing that large,
plausibly a reward-hacking interaction with the discrete pressure solve
that lowers the sensed pressure gradient while the wall does the work.
A banded wall actuation of just this kind is in fact
discernible in the developed-flow snapshots reported for memoryless
learnt channel controllers elsewhere, so the configuration is not
peculiar to the present solver, even if its precise coupling to the
discrete solver would need reproducing case by case before being read as
universal. The transferable statement is the one the Lorenz--96 surrogate
makes, that a memoryless policy off the right cadence either saturates or
fades. Stripes is fixed by construction, an imposed
open-loop forcing with no sensing, and is shown only as the reference for
what a static pattern costs energetically; the contrast here is that a
closed-loop learnt policy arrives at a box-scale pattern of its own.
GRU-MARL, carrying a per-patch hidden state, instead places its actuation
in register with the moving near-wall streaks of the detection-plane
field.

\subsection*{Removing reward-hacking pathways reveals genuine flow control}

\begin{figure}[htbp]
\centering
\includegraphics[width=\linewidth,height=0.40\textheight,keepaspectratio]{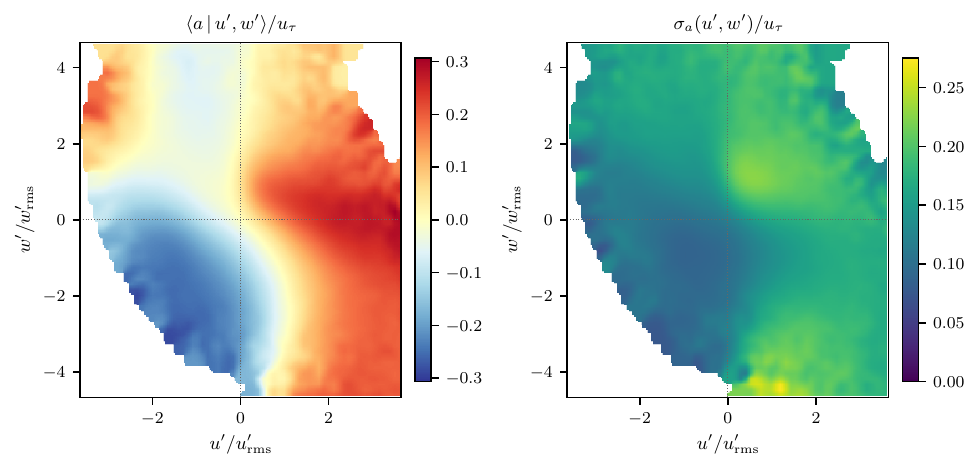}
\caption{GRU-MARL conditional policy response. Unlike the memoryless
vanilla-DRL actor of Fig.~\ref{fig:saturation}, whose entire policy is a
surface on the two sensed scalars, GRU-MARL has no two-dimensional input;
the panels show the action conditionally averaged over the rollout at
fixed patch-mean $(u',w')$ (left) and its spread (right). The spread is
the share of the action the local two scalars cannot account for,
supplied by the recurrence, the spatial ring and the hidden state.}
\label{fig:policymap}
\end{figure}

\begin{figure}[htbp]
\centering
\begin{tabular}{@{}c@{\hspace{8pt}}c@{}}
\includegraphics[width=0.47\textwidth,height=0.34\textheight,keepaspectratio]{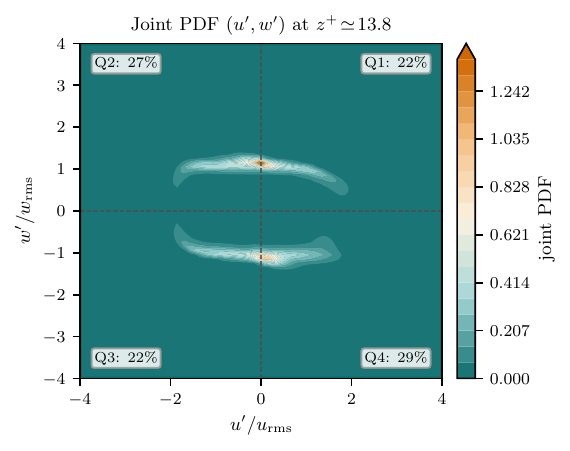} &
\includegraphics[width=0.47\textwidth,height=0.34\textheight,keepaspectratio]{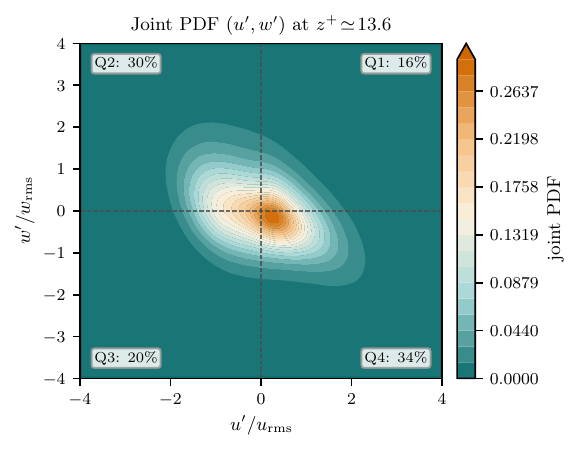} \\
{\footnotesize \textbf{a} vanilla DRL: $f(u',w')$} &
{\footnotesize \textbf{b} GRU-MARL: $f(u',w')$} \\[6pt]
\includegraphics[width=0.47\textwidth,height=0.34\textheight,keepaspectratio]{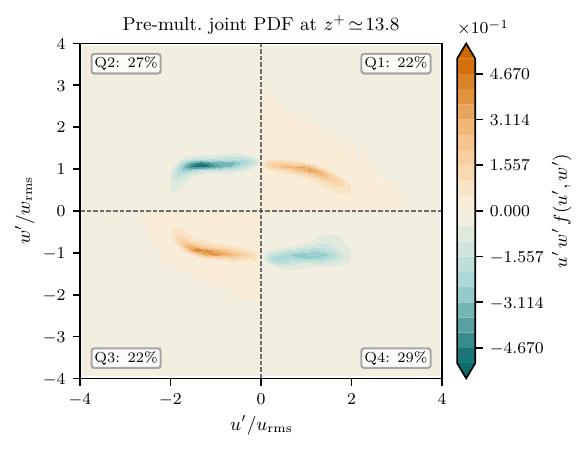} &
\includegraphics[width=0.47\textwidth,height=0.34\textheight,keepaspectratio]{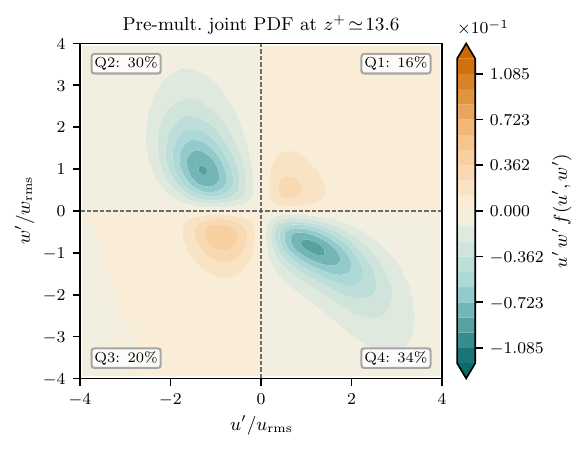} \\
{\footnotesize \textbf{c} vanilla DRL: $u'w'\,f(u',w')$} &
{\footnotesize \textbf{d} GRU-MARL: $u'w'\,f(u',w')$} \\
\end{tabular}
\caption{Detection-plane joint statistics of the streamwise and
wall-normal fluctuations: the raw density $f(u',w')$ (top row) and its
Reynolds-shear-weighted form $u'w'\,f(u',w')$ (bottom row), for vanilla
DRL (left) and GRU-MARL (right). All four are computed over the
statistically steady window, with the initial transient discarded so that
no start-up effect enters the statistics. GRU-MARL reshapes the second-
and fourth-quadrant sweep--ejection events that the memoryless switch
leaves broad.}
\label{fig:pdfs}
\end{figure}

Adding the recurrent core changes the character of the policy. GRU-MARL
reads a three-channel stencil over a $3\times3$ ring of patches together
with its own previous action, and carries a per-patch hidden state
through time. Its input is no longer the two scalars of the memoryless
policy, so the action surface of Fig.~\ref{fig:saturation} has no direct
analogue here; what can be drawn instead is the action conditionally
averaged over the rollout at fixed patch-mean $(u',w')$
(Fig.~\ref{fig:policymap}). At each point of that conditional map the
action keeps a spread the patch-mean coordinates cannot set, and that
residual variance is what the recurrence, the spatial ring and the hidden
state supply beyond the local instantaneous state. The effect on the
near-wall flow is read off the detection-plane joint statistics
(Fig.~\ref{fig:pdfs}): against the broad, switch-driven distribution left
by vanilla DRL, GRU-MARL reshapes the $(u',w')$ density and its
Reynolds-shear-weighted form in the second and fourth quadrants where the
sweep--ejection cycle carries the momentum that sets the friction.

\begin{figure}[htbp]
\centering
\includegraphics[width=0.82\textwidth,height=0.48\textheight,keepaspectratio]{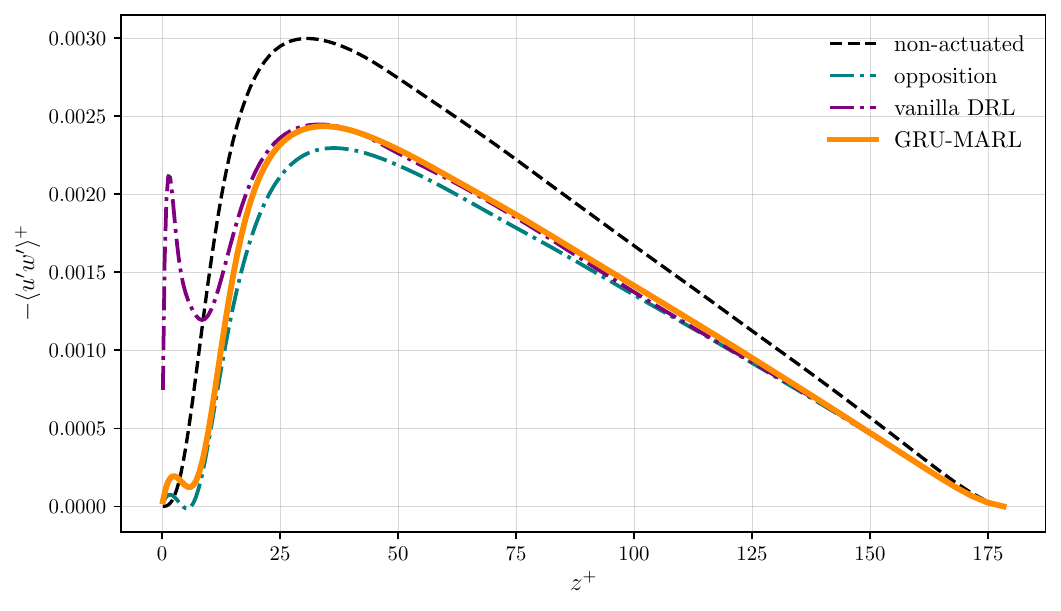}
\caption{Near-wall Reynolds shear stress $-\langle u'w'\rangle^{+}$ in
viscous units referenced to the uncontrolled friction velocity, pooled
over the snapshot ensemble, for the non-actuated channel, opposition,
vanilla DRL and GRU-MARL. GRU-MARL suppresses the momentum-carrying
shear stress close to opposition, while the memoryless switch leaves it
near the non-actuated level. The profiles are plotted at cell centres, so
the first point lies at $z^+=\Delta z^+(0)/2$ rather than at the wall;
$-\langle u'w'\rangle^{+}$ vanishes identically at $z=0$, where no-slip
holds on the wall-parallel components.}
\label{fig:profiles}
\end{figure}

The near-wall Reynolds shear stress confirms that the recurrent
controller acts on the flow rather than on the bookkeeping
(Fig.~\ref{fig:profiles}). Against the non-actuated reference, GRU-MARL
suppresses the $-\langle u'w'\rangle^{+}$ that carries momentum to the
wall, close to what opposition achieves, while the saturating
vanilla-DRL switch leaves it near the uncontrolled level and reorganises
the field at the energetic cost the budget has already exposed. The
recurrent policy reaches its operating point by a route that opposition
does not take. Both controllers carry a comparable linear coupling to the
streamwise velocity at the detection plane ($\rho_{\ww,u}\simeq0.46$ and
$0.47$), but GRU-MARL retains almost no coupling to the wall-normal
component, the anticorrelation that defines the opposition mechanism
having fallen from $-0.95$ to $+0.05$. It reaches an opposition-like
operating point without opposing, and with no opposition prior in the
architecture or the reward (the sensing correlations are reported in
Supplementary Table~S1, and the field correlation maps in Supplementary
Fig.~S1). That it lands close to opposition, $17.3\%$ against $21.4\%$ in
drag and $17.3\%$ against $21.5\%$ in dissipation, is the result rather
than a shortfall. The two reach comparable budgets by different means,
and GRU-MARL does so at two-fifths of the opposition peak amplitude. A
second route to genuine drag reduction therefore exists and can be
learnt, without the collapse onto degenerate policies that unguarded
objectives produce. A practical property follows from the parameter
sharing, and it is a property of the sharing rather than of the
multi-agent decomposition: a policy applied identically over a local
receptive field is translation-equivariant, so the same weights define
the policy on any grid. A network trained on a minimal flow
unit of $64\times64$ patches therefore transfers without change to the
$256\times256$ evaluation grid of a box sixteen times larger in
wall-parallel area, and every GRU-MARL number reported here is from that
transferred policy.

\subsection*{The credit-assignment and timescale failures are architectural}

Two of the three failures owe nothing to the fluid. The credit-assignment
collapse and the timescale collapse each reappear in the
fluid-dynamics-free surrogates introduced alongside them, which carry no
flow yet reproduce the saturation and the dead actuation seen at the wall,
marking them as properties of the control architecture rather than of
turbulence (Supplementary Figs.~S2 and~S3). The energy budget is
different in kind. The wall power the drag proxy omits is a physical
quantity with no analogue in a system that carries no actuation cost, so
the third fault is a property of the flow, one the channel exposes rather
than abstracts away. Construction, training and numerical checks for both
surrogates are given in the Supplementary Information.
 %%%%%%%%%%%%%%%%%%%%%%%%%%%%%%%%%%%
\section*{Discussion}

The results identify three distinct pathways through which reinforcement-learning controllers can appear successful while failing to improve the underlying physical objective: incomplete rewards, constraint-induced corruption of the learning signal, and partial observability of the controlled dynamics. In the present study these mechanisms are exposed in turbulent drag reduction, but they are not specific to fluid mechanics.

These mechanisms do not arise from deficiencies in reinforcement learning itself. Rather, they emerge when the learning problem provides only a partial description of the physical objective against which success is ultimately measured. In such circumstances, optimisation can improve the reported metric without improving the quantity that the control strategy is intended to tackle \citep{krakovna2020specification}.

In physical systems this problem becomes particularly acute because conservation laws, admissibility constraints and limited observations shape both the information available to the controller and the consequences of its actions. 
A controller may therefore improve the reported metric without improving the physical outcome, not because the learning algorithm is ineffective, but 
because the formulation of the control problem leaves important physical consequences unpenalised or unresolved.

Turbulent skin-friction drag reduction provides a particularly transparent setting in which to examine these effects because the governing equations are known, the conservation constraints are explicit, and the complete energy budget can be measured directly. This makes it possible to distinguish between apparent improvements in drag reduction and genuine reductions in total dissipation, and to isolate the separate roles played by reward definition, constraint handling and observability. 
The failure modes therefore arise from the formulation of the control problem rather than from the reinforcement-learning method itself.

The physical objective in wall-bounded turbulent drag reduction is well defined: reducing the drag exerted by the flow ultimately corresponds to lowering the total rate of energy dissipation in the system. In much of the recent reinforcement-learning literature, however, this objective has been replaced by a more readily measurable proxy, namely the reduction in pumping power inferred from the streamwise pressure gradient. While convenient, this quantity captures only one component of the overall energy balance and does not account for the work performed by the actuation itself on the flow \citep{guastoni2023drl,walchli2024minimalchannel}.

The need to distinguish nominal drag reduction from net energetic benefit is not new. Classical flow-control studies have long emphasised that any meaningful assessment of control performance must consider the net energy balance. For example, \citet{fukagata2009netpower} derived bounds on the net power savings achievable by actuated wall-bounded flows, while \citet{marusic2021energypath} argued that energy efficiency, rather than nominal drag reduction alone, provides the appropriate criterion for evaluating the practical value of a control strategy. A scheme that reduces the pressure gradient may therefore still be energetically unfavourable if the control action requires a comparable or larger energy input.

The three failure modes discussed below correspond to the three ways in which the control formulation can fail to represent the physical objective: through the reward, the constraint handling and the observations. Each is a property of a formulation the field has settled on rather than of the baseline we built to expose it. The parameter-shared arrangement with centralised training and a globally constrained action set is the template of recent multi-agent flow control \citep{guastoni2023drl,vignon2023rb,suarez2024marl3dcylinder,font2025separationbubble}; memoryless policies mapping an instantaneous local observation to an action are the norm in learnt channel control \citep{guastoni2023drl,zhou2025drlhighre}; and the amplitude-based actuation cost runs from \citet{kametani2011blowingsuction} through \citet{guastoni2023drl} to \citet{beneitez2025}.
%The mismatch between the physical objective and the quantity used to evaluate it lies at the heart of the three shortcomings discussed in this paper.

The first fault is structural. The mass-conservation constraint that makes blowing and suction admissible in a closed channel couples the actions of all agents through a mean-subtraction projection. When this projection is applied after the actor network rather than embedded within it, the policy-gradient update is computed for a control field different from the one actually imposed on the flow. This distorts the credit assignment mechanism on which learning relies. The remedy is straightforward and incurs no additional modelling cost: by implementing the projection as the final layer of the actor, its constant Jacobian becomes part of the computational graph, ensuring that gradients are taken with respect to the physically admissible actuation field.

The second fault concerns observability. The near-wall regeneration cycle evolves over
time scales of the order of one hundred viscous time units, substantially longer than the instantaneous flow snapshots presented to a memoryless policy. A static mapping from a single observation cannot reliably infer the phase of such a slowly evolving process. As a consequence, the optimisation tends to converge towards a saturated two-level actuation strategy, the familiar bang-bang regime that often emerges in continuous-control problems when rapid switching is left unpenalised \citep{seyde2021bangbang}.

To address this limitation, we introduce two complementary modifications. First, the actuation interval is selected to lie between the fast decorrelation time of the near-wall turbulence and the slower turnover time of the streak-regeneration cycle. Successive observations therefore contain meaningful dynamical information while still allowing multiple control updates within the lifetime of a coherent structure. Second, the policy is augmented with a recurrent architecture and an enlarged sensing stencil, providing the temporal memory and spatial context required to track streak-scale dynamics rather than reacting to the fastest fluctuations.

The third fault concerns the way energetic performance is evaluated and reported. The standard accounting framework assigns an actuation cost through a kinetic-energy-flux term that scales with the cube of the wall-normal velocity amplitude. At the amplitudes typically used in turbulent-flow control, this contribution is often two to three orders of magnitude smaller than the friction coefficient itself, causing the reported net-energy saving to become numerically indistinguishable from the drag reduction. More recent formulations replace this cubic estimate with a quadratic kinematic cost, sometimes applied only to blowing while effectively treating suction as energetically free \citep{hasegawa2011dissimilar}. Although different in form, both approaches share the same limitation: neither accounts for the signed pressure covariance through which a bounded-amplitude actuation exchanges mechanical power with the flow. One partial exception should be noted. \citet{zhou2025drlhighre}, following \citet{bewley2001predictive}, do include a wall-pressure term in their reported power input, but they take its absolute value inside the average, $\langle\lvert p'_w v_w\rvert\rangle$. A rectified magnitude is non-negative by construction, so it bounds the actuation cost without resolving the direction of the exchange, and it cannot enter the signed balance of~\eqref{eq:balance}. The distinction is decisive for the controllers examined here, whose failure is precisely that the wall does net positive work on the flow.

This omission becomes particularly important for reinforcement-learning controllers. Neither metric imposes a meaningful energetic penalty on bang-bang saturation beyond a largely amplitude-independent cost. As a result, a controller that spends most of its time at the actuation bounds can incur essentially the same reported energy expenditure as a smoother policy, despite performing substantially more work on the fluid. In the extreme case, a policy may converge to an actuation pattern whose length scale is set by the periodic domain rather than by the flow, continuously injecting energy while still appearing favourable under conventional drag-reduction metrics.

The energetic criterion itself is not new. That active control must be judged on net power rather than on drag alone has been standard in the flow-control literature since \citet{bewley2001predictive} and \citet{fukagata2009netpower}. What is new here is its use as a diagnostic: applied to learnt controllers, the same criterion separates policies that suppress turbulence from policies that pump the wall, and that separation is invisible to the metrics on which those policies are trained and reported. The physically relevant quantities are therefore the true wall-power input, represented by the pressure-covariance term in~\eqref{eq:powers}, and the corresponding net-energy balance. These are the measures advocated in the classical flow-control literature \citep{fukagata2009netpower,marusic2021energypath} and should likewise form the basis for evaluating learned controllers. Without them, a controller that actively pumps energy into the flow can appear indistinguishable from one that genuinely suppresses turbulent activity. The stripe and vanilla-DRL results reported in Table~\ref{tab:energy} provide a clear illustration of this ambiguity.

These considerations also suggest caution when comparing controllers trained and evaluated under different assumptions regarding domain size and energetic accounting. A controller evaluated in a minimal flow unit, where relaminarisation can substantially inflate the reported drag reduction, and whose actuation cost is estimated using a metric insensitive to bang-bang saturation, will naturally produce a larger headline figure than a controller assessed in a large computational domain using the true wall-power expenditure. The difference between these figures does not necessarily reflect a difference in control effectiveness, but may instead arise from the underlying evaluation methodology.

Viewed in this context, the $17\%$ drag reduction achieved by GRU-MARL should be interpreted as a conservative estimate obtained under more stringent evaluation conditions. The controller is assessed in a large domain, its energetic performance is measured using the physically relevant wall-power contribution, and the control formulation is designed to avoid known degeneracies associated with purely drag-based objectives.
%reward function is designed to avoid known degeneracies associated with purely drag-based objectives. 
We therefore view this result not as a final benchmark, but as a reference point for future comparisons performed under the same physical and energetic criteria.

All the results reported here are at $\Retau\simeq180$, and it is worth separating what should be expected to survive that restriction from what should not. The diagnosis should, since the three faults are properties of the control formulation rather than of the flow, and two of the three are reproduced with no fluid at all. The quantitative performance of the corrected controller, and the actuation cadence the near-wall cycle demands, are tied to this Reynolds number, and we do not claim that either carries over.

Several questions remain open. How the corrected policy reorganises the near-wall flow, and what aspects of the regeneration cycle it exploits beyond the instantaneous local state, is the subject of a companion study aimed at directly analysing the information content of the learned policy. Such an analysis may also provide new insight into the dynamics of the wall cycle itself. The present architecture is not fully optimised: the sensing stencil, recurrent memory and reward weighting all offer opportunities for further improvement while maintaining the same energetic standard. A natural next step is to determine whether the same corrections remain effective at higher Reynolds numbers, where the effectiveness of opposition-based control is known to diminish \citep{zhou2025drlhighre}.

Although demonstrated here in wall-bounded turbulence, the lesson is broader. Physical reinforcement-learning controllers, particularly with multi-agent architectures and decentralised training frameworks, should be evaluated against the full physical objective, with constraints included in the differentiable control loop and observations chosen to resolve the relevant dynamics. Without these conditions, improved reported performance may reflect exploitation of an incomplete formulation rather than genuine control of the system.

% =======================================================================
\section*{Methods}

\subsection*{Flow configuration}

The flow is an incompressible turbulent channel at $\Retau\simeq180$, the
lowest of the standard channel-flow reference simulations
\citep{moser1999channel} that reach $\Retau\simeq2000$
\citep{hoyas2006channelre2003}, in a half-channel of height $H$, periodic
in the wall-parallel directions,
with an actuated no-slip wall at the bottom and a symmetry condition at
the top, driven in constant-flow-rate mode so that the streamwise
pressure gradient $\avg{\dpdx}(t)$ adjusts to hold the bulk velocity at
$\Ubulk=1$. The uncontrolled mean gradient is
$\avg{\dpdx}_0\approx-4.10\times10^{-3}$ in bulk-velocity and
half-height units. Evaluation runs use a box of
$(L_x,L_y,H)=(10.68,3.2,1)$, that is
$(L_x^+,L_y^+,H^+)\simeq(1922,576,180)$, on a $256\times256\times100$
mesh with near-wall resolution $(\Delta x^+,\Delta y^+)\simeq(7.5,2.25)$
and a wall-normal grid clustered at the wall by a hyperbolic-tangent
stretching that places the first point inside $z^+\!\lesssim\!1$.
Actuation is a Dirichlet condition on the wall-normal velocity,
$w(x,y,0,t)=\ww(x,y,t)$, with the wall-parallel components no-slip and the
zero-net-mass constraint~\eqref{eq:zeromass} imposed at every actuation
step. The time step is set by a Courant limit of order one half and is of
order $10^{-3}$ bulk time units; each run is advanced for a spin-up of
$40$ bulk time units before any statistics are taken, and the
energy-budget terms of Table~\ref{tab:energy} are then averaged over a
window of order one hundred bulk time units in the statistically
stationary regime. The same uncontrolled initial condition seeds every
controlled run, so the comparison is at matched flow history.

\subsection*{Flow solver}

The Navier--Stokes equations are integrated with CaNS
\citep{costa2018cans}, a second-order finite-difference solver on a
staggered Cartesian grid that advances the momentum equations with a
low-storage third-order Runge--Kutta scheme and enforces incompressibility
by a pressure-correction step. The pressure Poisson equation is solved
with an FFT-based direct method in the two periodic directions and a
tridiagonal solve in the wall-normal direction. Constant-flow-rate
operation is imposed by adjusting the bulk pressure gradient each
substep. The simulations are run on GPUs using the accelerated CaNS
implementation \citep{costa2021gpucans} with the adaptive pencil
decomposition of cuDecomp \citep{romero2022cudecomp} for multi-GPU
communication. The wall actuation enters as a time-dependent Dirichlet
boundary condition updated at the actuation cadence.

\subsection*{Reinforcement-learning framework}

Control is provided through a multi-agent reinforcement-learning
framework that couples the flow solver to a PyTorch
\citep{paszke2019pytorch} policy and a PettingZoo
\citep{terry2020pettingzoo} parallel multi-agent environment
(Fig.~\ref{fig:overview}). The
environment advances CaNS as a subprocess for one actuation interval,
extracts the per-patch observations from the detection-plane fields,
passes them to the shared policy, and writes the resulting wall-velocity
field back as the boundary condition for the next interval; observations,
actions and rewards are exchanged through the solver's actuation
interface. Training follows the centralised-training,
decentralised-execution scheme established for multi-agent flow control
by \citet{vignon2023rb} and \citet{guastoni2023drl}: a single critic with
access to the joint state and action supplies the gradient, while at
deployment each patch acts on its local observation alone. Reinforcement
learning reached flow control through wake stabilisation
\citep{rabault2019cylinder} and was scaled by training across parallel
environments \citep{rabault2019multienv}, then carried to wings
\citep{vinuesa2022wings}, thermal convection
\citep{beintema2020rayleighbenard} and turbulence closure
\citep{novati2021turbmodel,bae2022wallmodel}, a line of work reviewed in
\citet{brunton2020mlreview}. In the channel it has been pursued with an
early network trained on a suboptimal target \citep{lee1997nncontrol},
single-agent convolutional controllers
\citep{han2020cnnchannel,lee2023drlchannel}, and studies of how the
Reynolds number changes the physics the policy exploits
\citep{varela2022actuators}, before the multi-agent formulation adopted
here. The framework is available at the repository in the Code
availability statement.

Every wall patch runs the same actor $\pi_\theta$, a convolutional
encoder feeding a single-layer GRU of width $d_h=64$ and a $\tanh$ head,
\begin{equation}
  o_{ij}\;\xrightarrow{\;\mathrm{CNN}\;}\;z_{ij}
  \;\xrightarrow{\;\mathrm{GRU}\;}\;(h_{ij}^{t+1},g_{ij}^{t})
  \;\xrightarrow{\;\mathrm{head}\;}\;
  a_{ij}^{t}=\tanh\!\bigl(W^{\top}\mathrm{LN}(g_{ij}^{t})+b\bigr),
  \label{eq:actor}
\end{equation}
where the per-patch hidden state $h_{ij}^{t}$ carries memory across
actuation steps and is what lets the policy follow the slow near-wall
cycle. The zero-mean projection~\eqref{eq:proj} is the actor's last
layer, so its Jacobian~\eqref{eq:jac} is differentiated through during
training; the resulting per-patch field is Gaussian-smoothed and
re-centred before it reaches the wall, which holds the net mass flux at
zero and prevents collapse to the discontinuous stripe pattern. The
centralised critic $Q_\phi$ reads the joint state and action and is used
during training only. Each patch evaluates the actor on its own local
observation ring, but the projection makes the emitted field depend on
the patch mean, so a single global reduction over the patch outputs is
required at every actuation step and the policy is not decentralised at
execution. The mean subtraction does not guarantee that the projected
action remains within the $\tanh$ bound; over the evaluation reported
here the largest projected action is $0.900$ of that bound, with no
attendant effect on the actuation amplitude or on the stability of the
coupled solver.

The policy is trained off-policy by multi-agent deterministic policy
gradient with the centralised critic \citep{lillicrap2015ddpg}, under an
energy-aware reward that offsets a drag-reduction term with penalties on
the action's temporal and spatial variation, its zero-mean residual and
its magnitude; the magnitude and smoothness penalties are the practical
form of the energy-aware objective, holding the policy off the saturation
bound that the wall power penalises. Training is on a minimal flow unit
\citep{jimenez1991mfu} of $(L_x^+,L_y^+)\simeq(481,144)$ for up to $500$
episodes of $1800$ actuation steps each, after which the shared
weights are deployed on the large evaluation box without retraining. The
network sizes, the learning update and the full set of hyperparameters
are given in the Supplementary Information (Supplementary Table~S2).

\subsection*{Wall units, observation and patches}

Throughout, the superscript $+$ denotes normalisation in viscous (wall)
units: lengths by $\nu/u_\tau$ and times by the viscous time
$\nu/u_\tau^{2}$, with friction velocity $u_\tau\simeq0.064$ and
$\nu=1/2870$ in bulk-velocity and half-height units, so that
$\Retau=u_\tau H/\nu\simeq180$. One viscous time unit is therefore
$\nu/u_\tau^{2}\simeq0.085$ bulk time units.

The wall is partitioned into square patches of $p=4$ grid cells, so the
$64\times64$ training grid carries $16\times16=256$ patches and the
$256\times256$ evaluation grid carries $64\times64=4096$; the same shared
weights are used in both. Each patch observes a $3\times3$ ring of
neighbouring patches with periodic closure, a three-channel $12\times12$
image of the plane-centred streamwise and wall-normal velocities at the
detection plane $z_d^+\simeq14$ together with its own previous action.
The two velocity channels are plane-mean-subtracted and scaled by
$\omega_{\max}=u_\tau$ so that the observation has an $\mathcal{O}(1)$
range. The patch sides are $\Delta x_p^+\simeq30$ and
$\Delta y_p^+\simeq9$, both below the spanwise streak spacing
$\lambda_y^+\simeq100$. What the policy senses, and where it is sensed,
shapes the law it learns \citep{paris2021sensor,vinuesa2024stateobservation},
which is the rationale for the widened detection stencil used here.

\subsection*{Actuation time scale}

The interval between two wall updates is a design parameter, and we fix
it from the time scale of the structures the controller must act on
rather than from the solver step. The near-wall streaks and the
regeneration cycle that maintains them, fed by the sweep and ejection
events that carry Reynolds shear stress across the buffer layer
\citep{wallace1972wallregion}, evolve over $\mathcal{O}(100)$ viscous
units \citep{jimenez1999autonomous,hamilton1995regeneration}. An
actuation interval close to a single viscous unit, $\Delta t_a^+\simeq1$,
forces the policy to act far below that scale, on near-uncorrelated
successive snapshots; with no memory to integrate, a policy driven that
fast collapses into the high-frequency switching that the vanilla-DRL
controller exhibits. We therefore set $\Delta t_a^+\simeq5$, equivalently
one update every $\simeq0.43$ bulk time units: short enough that several
updates fall within one streak lifetime, yet coarse enough that, paired
with the recurrent memory, the policy follows the streak-scale evolution
instead of chasing the fastest fluctuations, which is what gives GRU-MARL
its purchase on the near-wall cycle. The same effect can be isolated away
from the turbulence in a controlled two-timescale model, detailed in the
Supplementary Information (Supplementary Fig.~S3): with a clean
separation between a fast and a slow scale, an actuation interval at the
fast scale collapses the policy to saturated two-level switching and one
above the slow scale to near-zero actuation, leaving only an intermediate
window admissible. The interval is held fixed across training and
evaluation.

\subsection*{Energy budget and the definition of wall power}

The pumping power and wall power are defined in~\eqref{eq:powers}. The
viscous wall-traction work vanishes because continuity at the no-slip
wall forces $\partial_z w|_{\mathrm{wall}}=0$ pointwise, so the only
surviving surface-work term is the pressure covariance. The closed
balance is~\eqref{eq:balance}, the drag reduction is~\eqref{eq:dr}, and
the net gain is~\eqref{eq:netgain}. We report $\Ww$ rather than the
kinetic-energy-flux proxy~\eqref{eq:proxy} because the latter is a
function of the actuation amplitude only and therefore cannot detect
power delivered through the correlation of the wall velocity with the
wall pressure; a controller can keep $\Win$ small by bounding its output
while $\Ww$ remains large. The streamwise momentum balance
$\avg{\tauw}=\lvert\avg{\dpdx}\rvert H$ holds independently of the
actuation and is used as a per-run solver check; the same control volume
underlies the identity that splits the skin friction into the
contributions a controller can target \citep{fukagata2002fik}. Each controller is run
for of order one hundred bulk-convective times past a stationary
transient, and every term is measured directly over that window.

\subsection*{Reference and deliberately unguarded controllers}

Opposition applies $\ww=-(w_d-\overline{w_d})$, with $w_d$ the
wall-normal velocity at the detection plane $z_d^+\simeq15$ and the
plane-mean subtracted to meet~\eqref{eq:zeromass}; the rule is
saturation-free, its amplitude set by the sensed signal. It suppresses
drag by raising a virtual wall that screens the surface from the
near-wall vortices \citep{hammond1998oppositionmech} and carries across
channel and boundary-layer geometries at a comparable rate
\citep{stroh2015oppositioncomparison}. The open-loop stripe controller
imposes the fixed streamwise square wave~\eqref{eq:stripes} at the
saturation amplitude $w^\star$, a steady-pattern counterpart of the
open-loop travelling-wave and blowing-and-suction schemes that reach
large nominal drag reductions
\citep{quadrio2009stw,gatti2016stwhighre,kametani2011blowingsuction}. The vanilla-DRL controller is a
memoryless parameter-shared multi-agent policy that maps the pointwise
detection-plane pair to a scalar action $a=\tanh f_\theta(u',w')$ through
a small network, trained on the drag-reduction reward without the
projection layer, recurrence or energy term that define the corrected
controller; its learnt surface is
reproduced by an asymmetric $\tanh$ on $a_1 u'+a_2 w'$ with
$(a_1,a_2)\simeq(3.6,-1.6)$ to $R^2\simeq0.999$.

% =======================================================================
\section*{Data and code availability}

The modified CaNS flow solver, the multi-agent reinforcement-learning
training and evaluation framework, and the trained GRU-MARL policy are
available at \url{https://github.com/gmcavallazzi/CaNS_GRU-MARL}, so the
controller reported here can be re-run and evaluated.

% =======================================================================
\clearpage
\setcounter{figure}{0}
\setcounter{table}{0}
\setcounter{equation}{0}
\renewcommand{\thefigure}{S\arabic{figure}}
\renewcommand{\thetable}{S\arabic{table}}
\renewcommand{\theequation}{S\arabic{equation}}
\newcolumntype{P}[1]{>{\centering\arraybackslash}m{#1}}

\section*{Supplementary Information}
\noindent
This document reports the actuation--velocity coupling that supports the
main-text claim that GRU-MARL operates in a streamwise-sensing regime
distinct from the wall-normal opposition law, together with two
fluid-dynamics-free models that reproduce the credit-assignment and
actuation-interval pathologies from first principles, and it collects the
network architecture and the full training hyperparameters of the
controller. The flow quantities
are evaluated over the statistically
stationary window of the per-step wall and detection-plane fields, with
the spin-up transient discarded.

% =====================================================================
\section*{Actuation--velocity coupling}

Table~\ref{tab:sensing} reports the zero-lag Pearson correlations
between the wall-normal wall velocity $\ww$ and the detection-plane
fluctuations at $z_d^+\simeq15$. Opposition is the near-ideal
wall-normal opposer ($\rho_{\ww,w}\simeq-0.95$) and carries a moderate
streamwise coupling alongside it ($+0.46$); vanilla DRL locks onto $w$
with the opposite sign. GRU-MARL retains the same streamwise coupling as
opposition ($+0.47$) but loses the wall-normal anticorrelation almost
entirely ($+0.05$), so what distinguishes it is not the acquisition of
$u$-sensing but the absence of $w$-opposition.

\begin{table}[h]
\centering\small
\renewcommand{\arraystretch}{1.25}
\setlength{\tabcolsep}{14pt}
\begin{tabular}{@{}lrr@{}}
\toprule
controller & $\rho_{\ww,u}$ & $\rho_{\ww,w}$ \\
\midrule
opposition  & $+0.46$ & $\mathbf{-0.95}$ \\
vanilla DRL & $-0.08$ & $\mathbf{+0.96}$ \\
GRU-MARL    & $\mathbf{+0.47}$ & $+0.05$ \\
\bottomrule
\end{tabular}
\caption{Zero-lag Pearson correlations between the wall actuation and the
detection-plane velocity fluctuations at $z_d^+\simeq15$. Bold marks the
dominant linear coupling of each controller. Opposition opposes $w$ and
vanilla DRL follows $w$ with the opposite sign, while GRU-MARL has no
significant wall-normal coupling; its streamwise correlation is
comparable to that of opposition.}
\label{tab:sensing}
\end{table}

\begin{figure}[h]
\centering
\setlength{\tabcolsep}{2pt}
\renewcommand{\arraystretch}{1.0}
\begin{tabular}{@{}m{1.8em}@{}P{0.42\textwidth}@{\hspace{4pt}}P{0.42\textwidth}@{}}
 & $C_{\ww,u}(\Delta x^+,\Delta y^+)$ & $C_{\ww,w}(\Delta x^+,\Delta y^+)$ \\
\rotatebox[origin=c]{90}{\small opposition} &
\includegraphics[width=0.42\textwidth,height=0.24\textheight,keepaspectratio]{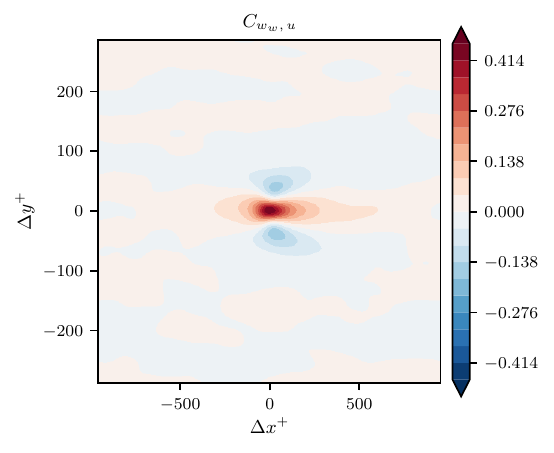} &
\includegraphics[width=0.42\textwidth,height=0.24\textheight,keepaspectratio]{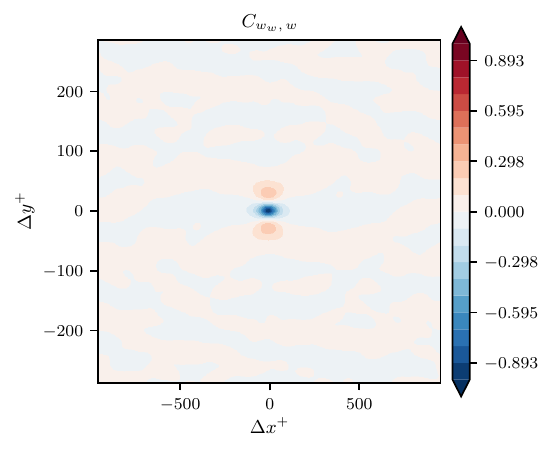} \\
\rotatebox[origin=c]{90}{\small vanilla DRL} &
\includegraphics[width=0.42\textwidth,height=0.24\textheight,keepaspectratio]{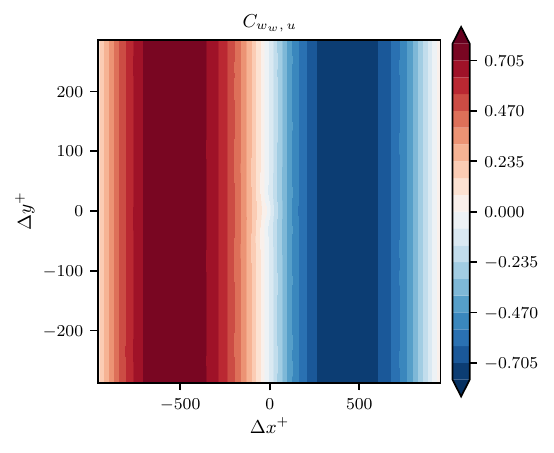} &
\includegraphics[width=0.42\textwidth,height=0.24\textheight,keepaspectratio]{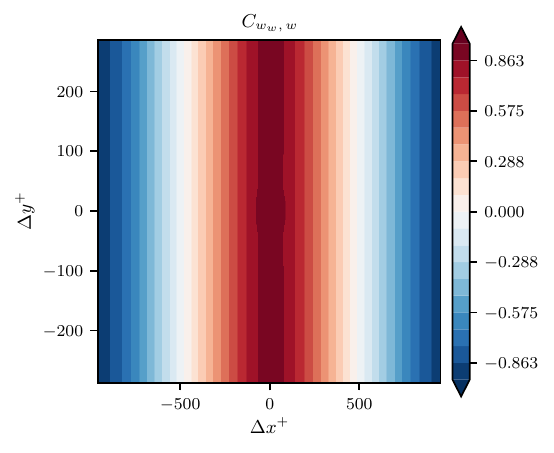} \\
\rotatebox[origin=c]{90}{\small GRU-MARL} &
\includegraphics[width=0.42\textwidth,height=0.24\textheight,keepaspectratio]{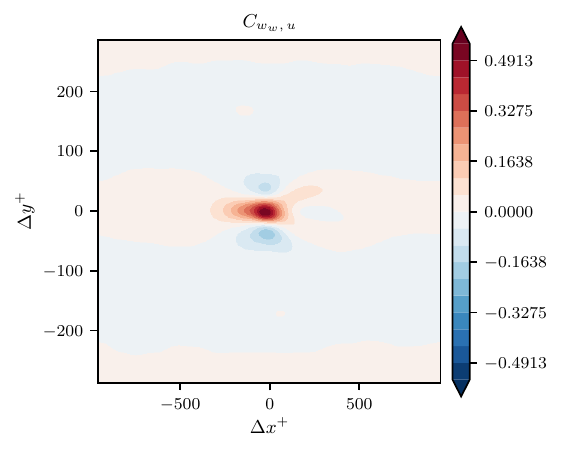} &
\includegraphics[width=0.42\textwidth,height=0.24\textheight,keepaspectratio]{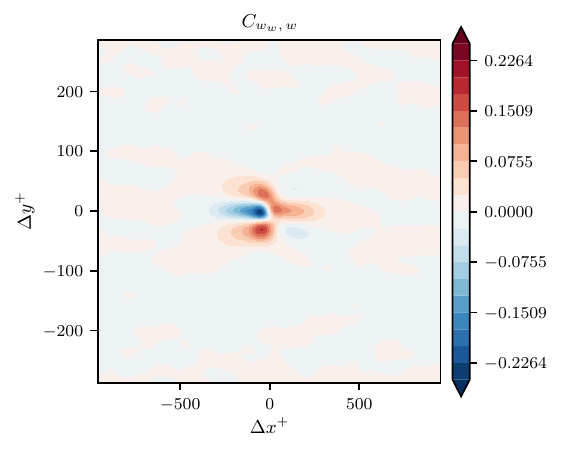} \\
\end{tabular}
\caption{Spatial cross-correlation maps between the wall actuation $\ww$
and the detection-plane streamwise (left) and wall-normal (right)
velocity, in viscous shifts, for opposition, vanilla DRL and GRU-MARL.
GRU-MARL couples to $u$ and not to $w$, the reverse of the wall-normal
opposer; vanilla DRL shows the imprint of its box-scale wave.}
\label{fig:corr}
\end{figure}

The full shift dependence of the coupling is shown in
Fig.~\ref{fig:corr}: the spatial cross-correlation maps
$C_{\ww,u}(\Delta x^+,\Delta y^+)$ and $C_{\ww,w}(\Delta x^+,\Delta y^+)$
between the actuation and the detection-plane velocities. The
GRU-MARL maps peak on the streamwise channel, the opposition maps on the
wall-normal channel, and the vanilla-DRL maps carry the periodic imprint
of its box-scale wave.

% =====================================================================
\section*{A minimal model of the credit-assignment pathology}

The credit-assignment pathology can be reproduced in a system with no
fluid dynamics. We set up $N=32$ agents, each emitting a bounded scalar
$a_i=\tanh(\cdot)$ from a single shared policy, a small multilayer
perceptron that reads a noisy local observation of a per-agent demand
$t_i$. A conserved resource forces the joint action onto its zero-mean
subspace, $a' = a - \tfrac1N\sum_j a_j$, before the quadratic reward
$r=-\langle(a'-t)^2\rangle$ is returned. The demand is redrawn every step
from a random latent context and carries a persistent mean $m_0$, which
stands in for a net wall actuation the conserved resource cannot supply.
Training uses the Adam optimiser on these random
demands, with noisy observations, so the learning curves fluctuate and
the optimum is approached rather than imposed. The two runs differ in one
respect only, the backward pass through the projection: with the
projection inside the actor the gradient carries its Jacobian
$\delta_{ij}-1/N$, while with the projection applied after the actor the
Jacobian is dropped, as it is when the zero-mean constraint is enforced
outside the network.

\begin{figure}[htbp]
\centering
\includegraphics[width=\linewidth,height=0.32\textheight,keepaspectratio]{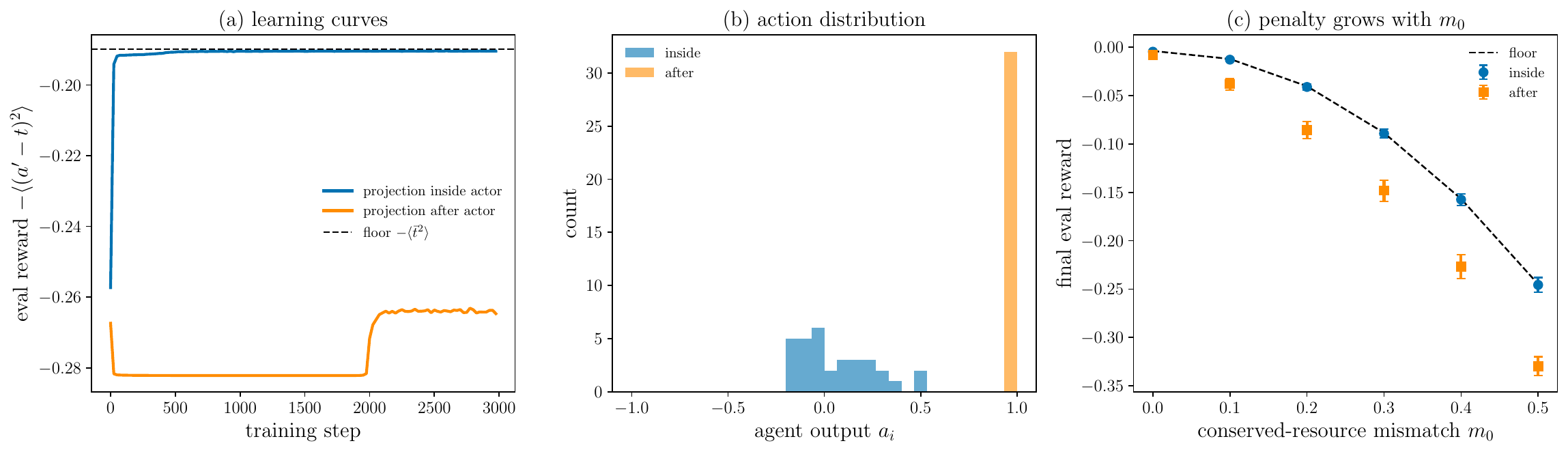}
\caption{Credit-assignment pathology in a fluid-dynamics-free model.
(a) Evaluation reward during training for the projection inside the actor
(blue) and applied after it (orange); the dashed line is the irreducible
floor $-\langle\bar t^{2}\rangle$ set by the conserved resource. (b)
Distribution of agent outputs $a_i$ at convergence: spread for the inside
gradient, piled against $\pm1$ for the omitted-Jacobian gradient. (c)
Final reward against the conserved-resource mismatch $m_0$ (mean and
spread over seeds); the inside policy tracks the floor, the after policy
falls increasingly below it as $m_0$ grows.}
\label{fig:credittoy}
\end{figure}

Figure~\ref{fig:credittoy} reports the outcome. With the Jacobian
retained, the policy learns down to the reward floor
$-\langle\bar t^{2}\rangle$ and no agent saturates (panel~a, blue;
panel~b). That floor is set by the conserved resource itself: a zero-mean
action cannot supply the mean of the demand, so a residual always
remains. With the Jacobian dropped, the same network and data settle well
above the floor, most agents pinned against their output bound (panel~a,
orange; panel~b). The dropped common-mode term acts as a steady forcing
that the bounded actor turns into saturation, the same failure the wall
controller shows but with no fluid present. Panel~(c) sweeps the mismatch
$m_0$: the inside policy stays on the floor across the range, while the
penalty from the dropped Jacobian grows with $m_0$ and closes as
$m_0\to0$. The collapse is thus a property of the shared conserved
resource and not of any dynamics, and putting the projection inside the
differentiated policy removes it.

% =====================================================================
\section*{A minimal two-timescale benchmark}

\begin{figure}[htbp]
\centering
\setlength{\tabcolsep}{3pt}
\renewcommand{\arraystretch}{1.0}
\begin{tabular}{@{}m{1.6em}@{}P{0.43\textwidth}@{\hspace{5pt}}P{0.43\textwidth}@{}}
 & {\small actuation $u(t)$} & {\small slow-variable energy} \\
\rotatebox[origin=c]{90}{\small fast $\Delta t_a$} &
\includegraphics[width=0.43\textwidth,height=0.18\textheight,keepaspectratio]{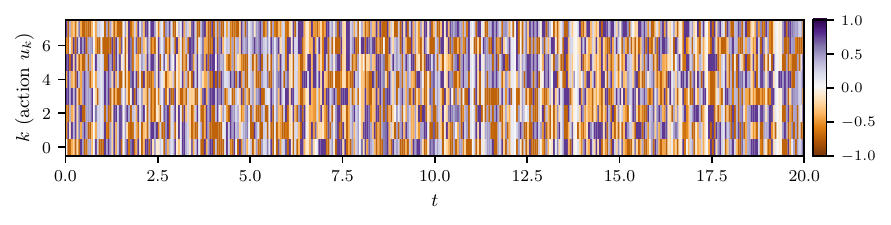} &
\includegraphics[width=0.43\textwidth,height=0.18\textheight,keepaspectratio]{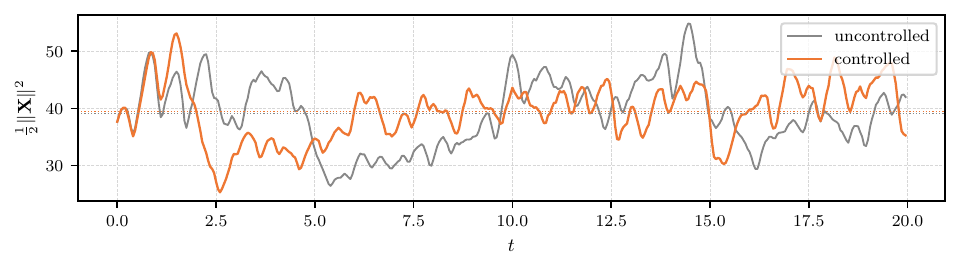} \\
\rotatebox[origin=c]{90}{\small good $\Delta t_a$} &
\includegraphics[width=0.43\textwidth,height=0.18\textheight,keepaspectratio]{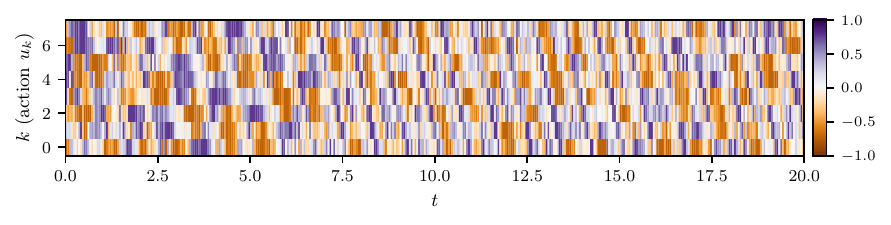} &
\includegraphics[width=0.43\textwidth,height=0.18\textheight,keepaspectratio]{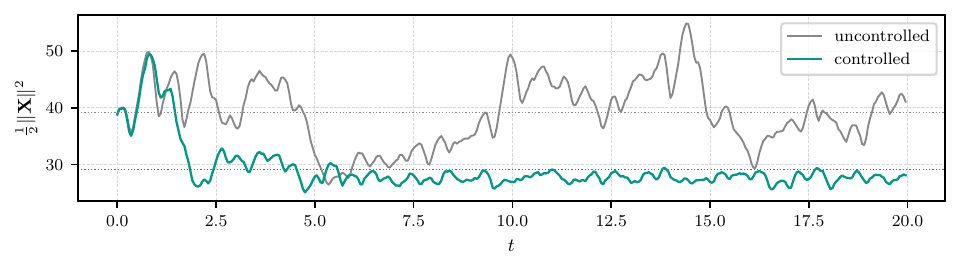} \\
\rotatebox[origin=c]{90}{\small slow $\Delta t_a$} &
\includegraphics[width=0.43\textwidth,height=0.18\textheight,keepaspectratio]{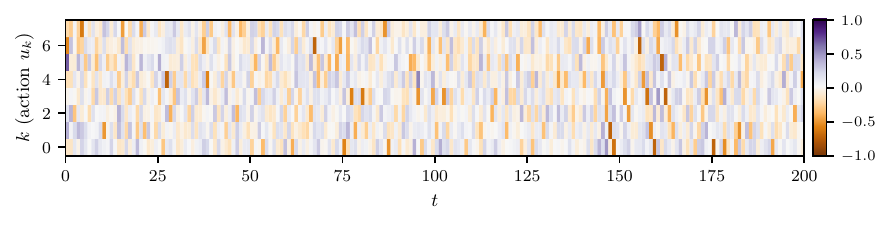} &
\includegraphics[width=0.43\textwidth,height=0.18\textheight,keepaspectratio]{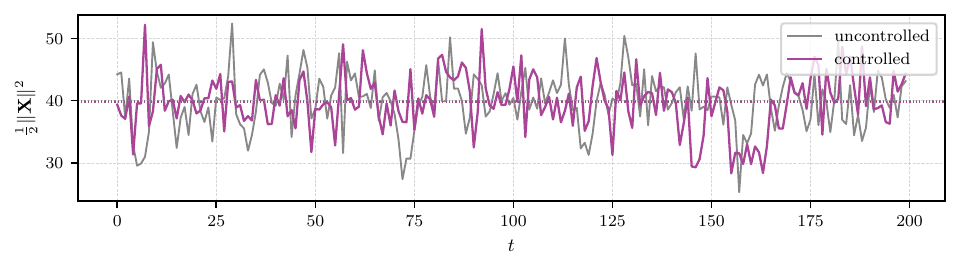} \\
\end{tabular}
\caption{Two-scale Lorenz--96 benchmark: actuation $u(t)$ (left) and
slow-variable energy (right) for three actuation intervals. A
fast-scale cadence collapses to saturated two-level switching (top); a
slow-scale cadence collapses to near-zero actuation on a stale
observation (bottom); only the intermediate cadence yields a structured,
energy-reducing control (middle).}
\label{fig:l96}
\end{figure}

The actuation interval can be studied away from the turbulence in a
controlled multiscale setting. We use a two-scale Lorenz--96 system
\citep{lorenz1996}, a standard surrogate for a flow with separated time
scales: a ring of slow variables $X_k$, each coupled to a bath of fast
variables $Y_{j,k}$ that relax on a time scale $1/c$ with $c=10$, so the
fast field decorrelates roughly every $0.1$ time units while the slow
variables keep memory for order one. Control enters the fast equation as
a per-site forcing under the same zero-mean projection across sites that
the wall actuation obeys, and the objective is to lower the
slow-variable energy. The only quantity varied between the runs of
Fig.~\ref{fig:l96} is the actuation interval $\Delta t_a$.

The two failure modes and the working window appear cleanly. When
$\Delta t_a$ is set near the fast decorrelation time, successive
observations are essentially uncorrelated and the policy, with nothing
coherent to integrate, collapses to a saturated two-level forcing that
switches sign at every step (Fig.~\ref{fig:l96}, top), the same degeneracy the
vanilla-DRL controller shows at the wall. When $\Delta t_a$ is set far
above the fast time, the observation is stale by the time the action is
applied and the policy collapses to near-zero actuation (bottom). Only an
intermediate cadence, short relative to the slow turnover yet long
relative to the fast decorrelation, produces a structured low-amplitude
forcing that lowers the slow-variable energy (middle). The channel
controller's choice $\Delta t_a^+\simeq5$, between the $\mathcal{O}(1)$
viscous fast scale and the $\mathcal{O}(100)$ streak lifetime, is the
wall-turbulence counterpart of that intermediate band.

% =====================================================================
\section*{GRU-MARL architecture and training details}

The shared actor reads each patch's three-channel $3\times3$ ring as an
image $o_{ij}\in\mathbb{R}^{3\times12\times12}$ and maps it to a scalar
action. A two-stage convolutional encoder, each stage
Conv($3\times3$)/GroupNorm($4$ groups)/ReLU/MaxPool($2$) with $[16,32]$
channels, reduces the ring to a $32\times3\times3$ map; a linear
projection to dimension $64$ feeds a single-layer GRU of width $64$ that
carries the per-patch hidden state; and a layer-norm with a $\tanh$ head,
initialised small so the policy begins near zero actuation, produces the
bounded scalar $a_{ij}^{t}\in[-1,1]$. The shared actor holds about
$4.9\times10^{4}$ parameters. The zero-mean projection is the actor's
last layer; the constant-per-patch field $A^{t}$ it returns is smoothed
and re-centred,
\begin{equation}
  \tilde A^{t}=G_\sigma * A^{t},\quad
  \tilde A^{t}\leftarrow\tilde A^{t}-\overline{\tilde A^{t}},\quad
  \ww = w^{\star}\,\tilde A^{t},
  \label{eq:pipeline}
\end{equation}
with $G_\sigma$ a periodic Gaussian of width $\sigma=1.5$ grid points, so
the field sent to the solver carries exactly zero net mass and does not
collapse to the discontinuous stripe pattern.

The centralised critic $Q_\phi$ acts on the joint field. The per-patch
rings are reassembled into the three-channel detection-plane state and
concatenated with the current action lifted to a constant-per-patch
field; the resulting four-channel stack passes through a three-stage
convolutional encoder ($[32,64,32]$ channels, Conv/GroupNorm/ReLU/MaxPool)
and a two-layer perceptron ($[256,128]$ with layer-norm and ReLU) to the
scalar value $Q_\phi(s,a)$. On the training grid it holds about
$6\times10^{5}$ parameters and is discarded at deployment, where each
patch runs the actor on its local ring alone.

Learning is by multi-agent deterministic policy gradient
\citep{lillicrap2015ddpg},
\begin{equation}
  \nabla_\theta J = \mathbb{E}_{(s,a)\sim\mathcal{D}}
  \bigl[\nabla_a Q_\phi(s,a)\big|_{a=\pi_\theta(s)}\,
        \nabla_\theta\pi_\theta(s)\bigr],
  \label{eq:dpg}
\end{equation}
with the differentiable projection's constant Jacobian carried in
$\nabla_\theta\pi_\theta$, and the centralised critic trained on the
temporal-difference residual
\begin{equation}
  \mathcal{L}_Q = \mathbb{E}\bigl[(Q_\phi(s,a)-r-\gamma\,
  Q_{\phi'}(s',\pi_{\theta'}(s')))^{2}\bigr].
  \label{eq:critic}
\end{equation}
The reward
\begin{equation}
  r_t = g\,\bigl(r_{\mathrm{DR}}
        -\lambda_1\mathcal{R}_{\mathrm{temp}}
        -\lambda_2\mathcal{R}_{\mathrm{spat}}
        -\lambda_3\mathcal{R}_{\mathrm{zero}}
        -\lambda_4\mathcal{R}_{\mathrm{mag}}\bigr)
  \label{eq:reward}
\end{equation}
combines a piecewise-linear drag term with penalties on the action
time-variation $\lVert a^t-a^{t-1}\rVert^2$, the in-plane total variation
of the wall field, the residual of the zero-mean constraint before
renormalisation, and the action magnitude $\lVert a^t\rVert^2$. The
magnitude penalty bounds the wall power directly, since
$\lvert\Ww\rvert\le(\langle\ww^2\rangle\langle p_w^2\rangle)^{1/2}$, while
the penalties on temporal and spatial variation suppress the coherent
actuation--pressure phase locking that would saturate that bound. The
wall power itself is measured for evaluation and does not enter the
reward. Both
networks are trained off-policy from a replay buffer with target copies
updated by a slow Polyak average; exploration adds bounded Gaussian noise
to the actions ahead of the projection layer, and because the recurrent
core carries state the transitions are replayed as short sequences so the
hidden state is rolled rather than reset at each update. Optimisation uses
Adam \citep{kingma2014adam}. Table~\ref{tab:hyper} lists the full set.

\begin{table}[h]
\centering\small
\renewcommand{\arraystretch}{1.2}
\setlength{\tabcolsep}{10pt}
\begin{tabular}{@{}ll@{}}
\toprule
Hyperparameter & Value \\
\midrule
\multicolumn{2}{@{}l}{\textit{Observation and action}}\\
Patch size                       & $4\times4$ grid cells \\
Observation ring                 & $3\times3$ patches, $3\times12\times12$ \\
Detection plane                  & $z_d^+\simeq14$ \\
Action interval $\Delta t_a^+$   & $5$ ($\simeq0.43$ bulk t.u.) \\
Saturation amplitude $w^{\star}$ & $0.0512$ ($=0.8\,u_\tau$) \\
Smoothing width $\sigma$         & $1.5$ grid points \\
\midrule
\multicolumn{2}{@{}l}{\textit{Networks}}\\
Actor encoder channels   & $[16,32]$ \\
GRU hidden size $d_h$     & $64$ \\
Actor parameters         & $\approx4.9\times10^{4}$ \\
Critic conv channels     & $[32,64,32]$ \\
Critic MLP widths        & $[256,128]$ \\
Critic parameters        & $\approx6\times10^{5}$ \\
\midrule
\multicolumn{2}{@{}l}{\textit{Optimisation (MADDPG)}}\\
Optimiser                      & Adam \\
Actor / critic learning rate   & $3\times10^{-5}$ / $5\times10^{-4}$ \\
Discount $\gamma$              & $0.98$ \\
Target smoothing $\tau$        & $2\times10^{-3}$ \\
Mini-batch                     & $128$ sequences \\
Sequence length / burn-in      & $16$ / $4$ steps \\
Gradient steps / interval      & $2$ every $20$ steps \\
Gradient-norm clip             & $0.5$ \\
Replay buffer                  & $15$ episodes \\
\midrule
\multicolumn{2}{@{}l}{\textit{Reward and schedule}}\\
Weights $(\lambda_1,\lambda_2,\lambda_3,\lambda_4)$ & $(0.3,0.25,2.0,0.3)$ \\
Reward gain $g$                & $10$ \\
Exploration noise $\sigma$     & $0.15\to0.01$ over $150$ ep. \\
Episodes $\times$ length       & $\le 500\times1800$ steps \\
\bottomrule
\end{tabular}
\caption{GRU-MARL architecture and training hyperparameters; quantities
in wall units where marked, with $u_\tau\simeq0.064$. Training is on the
minimal flow unit ($16\times16$ patches) and the same shared weights are
deployed on the $64\times64$ evaluation grid.}
\label{tab:hyper}
\end{table}

\bibliographystyle{plainnat}
\bibliography{references}

\begin{thebibliography}{62}
\providecommand{\natexlab}[1]{#1}
\providecommand{\url}[1]{\texttt{#1}}
\expandafter\ifx\csname urlstyle\endcsname\relax
  \providecommand{\doi}[1]{doi: #1}\else
  \providecommand{\doi}{doi: \begingroup \urlstyle{rm}\Url}\fi

\bibitem[Amodei et~al.(2016)Amodei, Olah, Steinhardt, Christiano, Schulman, and
  Man{\'e}]{amodei2016aisafety}
D.~Amodei, C.~Olah, J.~Steinhardt, P.~Christiano, J.~Schulman, and D.~Man{\'e}.
\newblock Concrete problems in {AI} safety.
\newblock \emph{arXiv preprint arXiv:1606.06565}, 2016.

\bibitem[Bae and Koumoutsakos(2022)]{bae2022wallmodel}
H.~J. Bae and P.~Koumoutsakos.
\newblock Scientific multi-agent reinforcement learning for wall-modelled
  turbulent flows.
\newblock \emph{Nature Communications}, 13\penalty0 (1):\penalty0 1--9, 2022.

\bibitem[Beintema et~al.(2020)Beintema, Corbetta, Biferale, and
  Toschi]{beintema2020rayleighbenard}
G.~Beintema, A.~Corbetta, L.~Biferale, and F.~Toschi.
\newblock Controlling rayleigh--b{\'e}nard convection via reinforcement
  learning.
\newblock \emph{Journal of Turbulence}, 21\penalty0 (9-10):\penalty0 585--605,
  2020.

\bibitem[Beneitez et~al.(2025)Beneitez, Cremades, Guastoni, and
  Vinuesa]{beneitez2025}
M.~Beneitez, A.~Cremades, L.~Guastoni, and R.~Vinuesa.
\newblock Improving turbulence control through explainable deep learning.
\newblock \emph{arXiv preprint arXiv:2504.02354}, 2025.

\bibitem[Bewley et~al.(2001)Bewley, Moin, and Temam]{bewley2001predictive}
T.~R. Bewley, P.~Moin, and R.~Temam.
\newblock {DNS}-based predictive control of turbulence: an optimal benchmark
  for feedback algorithms.
\newblock \emph{Journal of Fluid Mechanics}, 447:\penalty0 179--225, 2001.

\bibitem[Brunton et~al.(2020)Brunton, Noack, and
  Koumoutsakos]{brunton2020mlreview}
S.~L. Brunton, B.~R. Noack, and P.~Koumoutsakos.
\newblock Machine learning for fluid mechanics.
\newblock \emph{Annual Review of Fluid Mechanics}, 52:\penalty0 477--508, 2020.

\bibitem[Cavallazzi et~al.(2025)Cavallazzi, Guastoni, Vinuesa, and
  Pinelli]{cavallazzi2024wallcycle}
G.~M. Cavallazzi, L.~Guastoni, R.~Vinuesa, and A.~Pinelli.
\newblock Deep reinforcement learning for the management of the wall
  regeneration cycle in wall-bounded turbulent flows.
\newblock \emph{Flow, Turbulence and Combustion}, 115:\penalty0 1291--1317,
  2025.
\newblock \doi{10.1007/s10494-024-00609-4}.

\bibitem[Choi et~al.(1994)Choi, Moin, and Kim]{choi1994opposition}
H.~Choi, P.~Moin, and J.~Kim.
\newblock Active turbulence control for drag reduction in wall-bounded flows.
\newblock \emph{Journal of Fluid Mechanics}, 262:\penalty0 75--110, 1994.

\bibitem[Costa(2018)]{costa2018cans}
P.~Costa.
\newblock A {FFT}-based finite-difference solver for massively-parallel direct
  numerical simulations of turbulent flows.
\newblock \emph{Computers \& Mathematics with Applications}, 76\penalty0
  (8):\penalty0 1853--1862, 2018.

\bibitem[Costa et~al.(2021)Costa, Phillips, Brandt, and
  Fatica]{costa2021gpucans}
P.~Costa, E.~Phillips, L.~Brandt, and M.~Fatica.
\newblock {GPU} acceleration of {CaNS} for massively-parallel direct numerical
  simulations of canonical fluid flows.
\newblock \emph{Computers \& Mathematics with Applications}, 81:\penalty0
  502--511, 2021.

\bibitem[Foerster et~al.(2018)Foerster, Farquhar, Afouras, Nardelli, and
  Whiteson]{foerster2018coma}
J.~N. Foerster, G.~Farquhar, T.~Afouras, N.~Nardelli, and S.~Whiteson.
\newblock Counterfactual multi-agent policy gradients.
\newblock In \emph{Proceedings of the AAAI Conference on Artificial
  Intelligence}, volume~32, 2018.

\bibitem[Font et~al.(2025)Font, Alc{\'a}ntara-{\'A}vila, Rabault, Vinuesa, and
  Lehmkuhl]{font2025separationbubble}
B.~Font, F.~Alc{\'a}ntara-{\'A}vila, J.~Rabault, R.~Vinuesa, and O.~Lehmkuhl.
\newblock Deep reinforcement learning for active flow control in a turbulent
  separation bubble.
\newblock \emph{Nature Communications}, 16\penalty0 (1):\penalty0 1422, 2025.

\bibitem[Fukagata et~al.(2002)Fukagata, Iwamoto, and Kasagi]{fukagata2002fik}
K.~Fukagata, K.~Iwamoto, and N.~Kasagi.
\newblock Contribution of {R}eynolds stress distribution to the skin friction
  in wall-bounded flows.
\newblock \emph{Physics of Fluids}, 14\penalty0 (11):\penalty0 L73--L76, 2002.

\bibitem[Fukagata et~al.(2009)Fukagata, Sugiyama, and
  Kasagi]{fukagata2009netpower}
K.~Fukagata, K.~Sugiyama, and N.~Kasagi.
\newblock On the lower bound of net driving power in controlled duct flows.
\newblock \emph{Physica D}, 238\penalty0 (13):\penalty0 1082--1086, 2009.

\bibitem[Garnier et~al.(2021)Garnier, Viquerat, Rabault, Larcher, Kuhnle, and
  Hachem]{garnier2021reviewdrl}
P.~Garnier, J.~Viquerat, J.~Rabault, A.~Larcher, A.~Kuhnle, and E.~Hachem.
\newblock A review on deep reinforcement learning for fluid mechanics.
\newblock \emph{Computers \& Fluids}, 225:\penalty0 104973, 2021.
\newblock \doi{10.1016/j.compfluid.2021.104973}.

\bibitem[Gatti and Quadrio(2016)]{gatti2016stwhighre}
D.~Gatti and M.~Quadrio.
\newblock {R}eynolds-number dependence of turbulent skin-friction drag
  reduction induced by spanwise forcing.
\newblock \emph{Journal of Fluid Mechanics}, 802:\penalty0 553--582, 2016.

\bibitem[Guastoni et~al.(2023)Guastoni, Rabault, Schlatter, Azizpour, and
  Vinuesa]{guastoni2023drl}
L.~Guastoni, J.~Rabault, P.~Schlatter, H.~Azizpour, and R.~Vinuesa.
\newblock Deep reinforcement learning for turbulent drag reduction in channel
  flows.
\newblock \emph{The European Physical Journal E}, 46\penalty0 (4):\penalty0 27,
  2023.

\bibitem[Hamilton et~al.(1995)Hamilton, Kim, and
  Waleffe]{hamilton1995regeneration}
J.~M. Hamilton, J.~Kim, and F.~Waleffe.
\newblock Regeneration mechanisms of near-wall turbulence structures.
\newblock \emph{Journal of Fluid Mechanics}, 287:\penalty0 317--348, 1995.

\bibitem[Hammond et~al.(1998)Hammond, Bewley, and
  Moin]{hammond1998oppositionmech}
E.~P. Hammond, T.~R. Bewley, and P.~Moin.
\newblock Observed mechanisms for turbulence attenuation and enhancement in
  opposition-controlled wall-bounded flows.
\newblock \emph{Physics of Fluids}, 10\penalty0 (9):\penalty0 2421--2423, 1998.

\bibitem[Han and Huang(2020)]{han2020cnnchannel}
B.-Z. Han and W.-X. Huang.
\newblock Active control for drag reduction of turbulent channel flow based on
  convolutional neural networks.
\newblock \emph{Physics of Fluids}, 32\penalty0 (9):\penalty0 095108, 2020.

\bibitem[Hasegawa and Kasagi(2011)]{hasegawa2011dissimilar}
Y.~Hasegawa and N.~Kasagi.
\newblock Dissimilar control of momentum and heat transfer in a fully developed
  turbulent channel flow.
\newblock \emph{Journal of Fluid Mechanics}, 683:\penalty0 57--93, 2011.

\bibitem[Hausknecht and Stone(2015)]{hausknecht2015drqn}
M.~Hausknecht and P.~Stone.
\newblock Deep recurrent {Q}-learning for partially observable {MDPs}.
\newblock In \emph{AAAI Fall Symposium on Sequential Decision Making for
  Intelligent Agents}, 2015.

\bibitem[Hoyas and Jim{\'e}nez(2006)]{hoyas2006channelre2003}
S.~Hoyas and J.~Jim{\'e}nez.
\newblock Scaling of the velocity fluctuations in turbulent channels up to
  {$Re_\tau = 2003$}.
\newblock \emph{Physics of Fluids}, 18:\penalty0 011702, 2006.

\bibitem[Jim{\'e}nez and Moin(1991)]{jimenez1991mfu}
J.~Jim{\'e}nez and P.~Moin.
\newblock The minimal flow unit in near-wall turbulence.
\newblock \emph{Journal of Fluid Mechanics}, 225:\penalty0 213--240, 1991.

\bibitem[Jim{\'e}nez and Pinelli(1999)]{jimenez1999autonomous}
J.~Jim{\'e}nez and A.~Pinelli.
\newblock The autonomous cycle of near-wall turbulence.
\newblock \emph{Journal of Fluid Mechanics}, 389:\penalty0 335--359, 1999.

\bibitem[Kaelbling et~al.(1998)Kaelbling, Littman, and
  Cassandra]{kaelbling1998pomdp}
L.~P. Kaelbling, M.~L. Littman, and A.~R. Cassandra.
\newblock Planning and acting in partially observable stochastic domains.
\newblock \emph{Artificial Intelligence}, 101\penalty0 (1--2):\penalty0
  99--134, 1998.

\bibitem[Kametani and Fukagata(2011)]{kametani2011blowingsuction}
Y.~Kametani and K.~Fukagata.
\newblock Direct numerical simulation of spatially developing turbulent
  boundary layers with uniform blowing or suction.
\newblock \emph{Journal of Fluid Mechanics}, 681:\penalty0 154--172, 2011.

\bibitem[Kametani et~al.(2015)Kametani, Fukagata, {\"O}rl{\"u}, and
  Schlatter]{kametani2015blowingsuctionTBL}
Y.~Kametani, K.~Fukagata, R.~{\"O}rl{\"u}, and P.~Schlatter.
\newblock Effect of uniform blowing/suction in a turbulent boundary layer at
  moderate {R}eynolds number.
\newblock \emph{International Journal of Heat and Fluid Flow}, 55:\penalty0
  132--142, 2015.

\bibitem[Kingma and Ba(2014)]{kingma2014adam}
D.~P. Kingma and J.~Ba.
\newblock {A}dam: a method for stochastic optimization.
\newblock \emph{arXiv preprint arXiv:1412.6980}, 2014.

\bibitem[Kline et~al.(1967)Kline, Reynolds, Schraub, and
  Runstadler]{kline1967structure}
S.~J. Kline, W.~C. Reynolds, F.~A. Schraub, and P.~W. Runstadler.
\newblock The structure of turbulent boundary layers.
\newblock \emph{Journal of Fluid Mechanics}, 30\penalty0 (4):\penalty0
  741--773, 1967.

\bibitem[Krakovna et~al.(2020)Krakovna, Uesato, Mikulik, Rahtz, Everitt, Kumar,
  Kenton, Leike, and Legg]{krakovna2020specification}
V.~Krakovna, J.~Uesato, V.~Mikulik, M.~Rahtz, T.~Everitt, R.~Kumar, Z.~Kenton,
  J.~Leike, and S.~Legg.
\newblock Specification gaming: the flip side of {AI} ingenuity.
\newblock \emph{DeepMind Blog}, 3:\penalty0 40--53, 2020.

\bibitem[Lee et~al.(1997)Lee, Kim, Babcock, and Goodman]{lee1997nncontrol}
C.~Lee, J.~Kim, D.~Babcock, and R.~Goodman.
\newblock Application of neural networks to turbulence control for drag
  reduction.
\newblock \emph{Physics of Fluids}, 9\penalty0 (6):\penalty0 1740--1747, 1997.

\bibitem[Lee et~al.(1998)Lee, Kim, and Choi]{lee1998suboptimal}
C.~Lee, J.~Kim, and H.~Choi.
\newblock Suboptimal control of turbulent channel flow for drag reduction.
\newblock \emph{Journal of Fluid Mechanics}, 358:\penalty0 245--258, 1998.

\bibitem[Lee et~al.(2023)Lee, Kim, and Lee]{lee2023drlchannel}
T.~Lee, J.~Kim, and C.~Lee.
\newblock Turbulence control for drag reduction through deep reinforcement
  learning.
\newblock \emph{Physical Review Fluids}, 8\penalty0 (2):\penalty0 024604, 2023.

\bibitem[Lillicrap et~al.(2015)Lillicrap, Hunt, Pritzel, Heess, Erez, Tassa,
  Silver, and Wierstra]{lillicrap2015ddpg}
T.~P. Lillicrap, J.~J. Hunt, A.~Pritzel, N.~Heess, T.~Erez, Y.~Tassa,
  D.~Silver, and D.~Wierstra.
\newblock Continuous control with deep reinforcement learning.
\newblock \emph{arXiv preprint arXiv:1509.02971}, 2015.

\bibitem[Lorenz(1996)]{lorenz1996}
E.~N. Lorenz.
\newblock Predictability: a problem partly solved.
\newblock In \emph{Proc.\ Seminar on Predictability}, volume~1, pages 1--18.
  ECMWF, Reading, UK, 1996.

\bibitem[Marusic et~al.(2021)Marusic, Chandran, Rouhi, Fu, Wine, Holloway,
  Chung, and Smits]{marusic2021energypath}
I.~Marusic, D.~Chandran, A.~Rouhi, M.~K. Fu, D.~Wine, B.~Holloway, D.~Chung,
  and A.~J. Smits.
\newblock An energy-efficient pathway to turbulent drag reduction.
\newblock \emph{Nature Communications}, 12:\penalty0 5805, 2021.

\bibitem[Moser et~al.(1999)Moser, Kim, and Mansour]{moser1999channel}
R.~D. Moser, J.~Kim, and N.~N. Mansour.
\newblock Direct numerical simulation of turbulent channel flow up to {$Re_\tau
  = 590$}.
\newblock \emph{Physics of Fluids}, 11\penalty0 (4):\penalty0 943--945, 1999.

\bibitem[Novati et~al.(2021)Novati, de~Laroussilhe, and
  Koumoutsakos]{novati2021turbmodel}
G.~Novati, H.~L. de~Laroussilhe, and P.~Koumoutsakos.
\newblock Automating turbulence modeling by multi-agent reinforcement learning.
\newblock \emph{Nature Machine Intelligence}, 3:\penalty0 87--96, 2021.

\bibitem[Paris et~al.(2021)Paris, Beneddine, and Dandois]{paris2021sensor}
R.~Paris, S.~Beneddine, and J.~Dandois.
\newblock Robust flow control and optimal sensor placement using deep
  reinforcement learning.
\newblock \emph{Journal of Fluid Mechanics}, 913, 2021.

\bibitem[Paszke et~al.(2019)Paszke, Gross, Massa, Lerer, Bradbury, Chanan,
  Killeen, Lin, Gimelshein, Antiga, et~al.]{paszke2019pytorch}
A.~Paszke, S.~Gross, F.~Massa, A.~Lerer, J.~Bradbury, G.~Chanan, T.~Killeen,
  Z.~Lin, N.~Gimelshein, L.~Antiga, et~al.
\newblock {P}y{T}orch: an imperative style, high-performance deep learning
  library.
\newblock In \emph{Advances in Neural Information Processing Systems 32}, 2019.

\bibitem[Quadrio et~al.(2009)Quadrio, Ricco, and Viotti]{quadrio2009stw}
M.~Quadrio, P.~Ricco, and C.~Viotti.
\newblock Streamwise-travelling waves of spanwise wall velocity for turbulent
  drag reduction.
\newblock \emph{Journal of Fluid Mechanics}, 627:\penalty0 161--178, 2009.

\bibitem[Rabault and Kuhnle(2019)]{rabault2019multienv}
J.~Rabault and A.~Kuhnle.
\newblock Accelerating deep reinforcement learning strategies of flow control
  through a multi-environment approach.
\newblock \emph{Physics of Fluids}, 31\penalty0 (9):\penalty0 094105, 2019.

\bibitem[Rabault et~al.(2019)Rabault, Kuchta, Jensen, R{\'e}glade, and
  Cerardi]{rabault2019cylinder}
J.~Rabault, M.~Kuchta, A.~Jensen, U.~R{\'e}glade, and N.~Cerardi.
\newblock Artificial neural networks trained through deep reinforcement
  learning discover control strategies for active flow control.
\newblock \emph{Journal of Fluid Mechanics}, 865:\penalty0 281--302, 2019.

\bibitem[Romero et~al.(2022)Romero, Costa, and Fatica]{romero2022cudecomp}
J.~Romero, P.~Costa, and M.~Fatica.
\newblock Distributed-memory simulations of turbulent flows on modern {GPU}
  systems using an adaptive pencil decomposition library.
\newblock In \emph{Proceedings of the Platform for Advanced Scientific
  Computing Conference (PASC '22)}, 2022.

\bibitem[Seyde et~al.(2021)Seyde, Gilitschenski, Schwarting, Stellato,
  Riedmiller, Wulfmeier, and Rus]{seyde2021bangbang}
T.~Seyde, I.~Gilitschenski, W.~Schwarting, B.~Stellato, M.~Riedmiller,
  M.~Wulfmeier, and D.~Rus.
\newblock Is bang-bang control all you need? {S}olving continuous control with
  {B}ernoulli policies.
\newblock In \emph{Advances in Neural Information Processing Systems},
  volume~34, pages 27209--27221, 2021.

\bibitem[Skalse et~al.(2022)Skalse, Howe, Krasheninnikov, and
  Krueger]{skalse2022hacking}
J.~Skalse, N.~H.~R. Howe, D.~Krasheninnikov, and D.~Krueger.
\newblock Defining and characterizing reward hacking.
\newblock In \emph{Advances in Neural Information Processing Systems},
  volume~35, pages 9460--9471, 2022.

\bibitem[Sonoda et~al.(2023)Sonoda, Liu, Itoh, and Hasegawa]{sonoda2023rl}
T.~Sonoda, Z.~Liu, T.~Itoh, and Y.~Hasegawa.
\newblock Reinforcement learning of control strategies for reducing skin
  friction drag in a fully developed turbulent channel flow.
\newblock \emph{Journal of Fluid Mechanics}, 960:\penalty0 A30, 2023.
\newblock \doi{10.1017/jfm.2023.147}.

\bibitem[Spalart and McLean(2011)]{spalart2011dragreduction}
P.~R. Spalart and J.~D. McLean.
\newblock Drag reduction: enticing turbulence, and then an industry.
\newblock \emph{Philosophical Transactions of the Royal Society A},
  369\penalty0 (1940):\penalty0 1556--1569, 2011.

\bibitem[Stroh et~al.(2015)Stroh, Frohnapfel, Schlatter, and
  Hasegawa]{stroh2015oppositioncomparison}
A.~Stroh, B.~Frohnapfel, P.~Schlatter, and Y.~Hasegawa.
\newblock A comparison of opposition control in turbulent boundary layer and
  turbulent channel flow.
\newblock \emph{Physics of Fluids}, 27\penalty0 (7):\penalty0 075101, 2015.

\bibitem[Su{\'a}rez et~al.(2024)Su{\'a}rez, Alc{\'a}ntara-{\'A}vila, Francisco,
  Jean, Arnau, Bernat, Lehmkuhl, and Vinuesa]{suarez2024marl3dcylinder}
P.~Su{\'a}rez, F.~Alc{\'a}ntara-{\'A}vila, A.~Francisco, R.~Jean, R.~Arnau,
  F.~Bernat, O.~Lehmkuhl, and R.~Vinuesa.
\newblock Flow control of three-dimensional cylinders transitioning to
  turbulence via multi-agent reinforcement learning, 2024.
\newblock arXiv:2405.17210.

\bibitem[Sutton and Barto(2018)]{sutton2018rl}
R.~S. Sutton and A.~G. Barto.
\newblock \emph{Reinforcement learning: an introduction}.
\newblock MIT Press, 2nd edition, 2018.

\bibitem[Terry et~al.(2020)Terry, Black, Hari, Santos, Dieffendahl, Williams,
  Lokesh, Horsch, and Ravi]{terry2020pettingzoo}
J.~K. Terry, B.~Black, A.~Hari, L.~S. Santos, C.~Dieffendahl, N.~L. Williams,
  C.~Lokesh, M.~Horsch, and P.~Ravi.
\newblock {PettingZoo}: Gym for multi-agent reinforcement learning.
\newblock \emph{arXiv preprint arXiv:2009.14471}, 2020.

\bibitem[Varela et~al.(2022)Varela, Su{\'a}rez, Alc{\'a}ntara-{\'A}vila,
  Mir{\'o}, Rabault, Font, Garc{\'i}a-Cuevas, Lehmkuhl, and
  Vinuesa]{varela2022actuators}
P.~Varela, P.~Su{\'a}rez, F.~Alc{\'a}ntara-{\'A}vila, A.~Mir{\'o}, J.~Rabault,
  B.~Font, L.~M. Garc{\'i}a-Cuevas, O.~Lehmkuhl, and R.~Vinuesa.
\newblock Deep reinforcement learning for flow control exploits different
  physics for increasing {R}eynolds number regimes.
\newblock \emph{Actuators}, 11\penalty0 (12), 2022.
\newblock \doi{10.3390/act11120359}.

\bibitem[Vignon et~al.(2023{\natexlab{a}})Vignon, Rabault, Vasanth,
  Alc{\'a}ntara-{\'A}vila, Mortensen, and Vinuesa]{vignon2023rb}
C.~Vignon, J.~Rabault, J.~Vasanth, F.~Alc{\'a}ntara-{\'A}vila, M.~Mortensen,
  and R.~Vinuesa.
\newblock Effective control of two-dimensional {R}ayleigh--{B}{\'e}nard
  convection: invariant multi-agent reinforcement learning is all you need.
\newblock \emph{Physics of Fluids}, 35\penalty0 (6):\penalty0 065146,
  2023{\natexlab{a}}.

\bibitem[Vignon et~al.(2023{\natexlab{b}})Vignon, Rabault, and
  Vinuesa]{vignon2023review}
C.~Vignon, J.~Rabault, and R.~Vinuesa.
\newblock Recent advances in applying deep reinforcement learning for flow
  control: perspectives and future directions.
\newblock \emph{Physics of Fluids}, 35\penalty0 (3):\penalty0 031301,
  2023{\natexlab{b}}.

\bibitem[Vinuesa et~al.(2022)Vinuesa, Lehmkuhl, Lozano-Dur{\'a}n, and
  Rabault]{vinuesa2022wings}
R.~Vinuesa, O.~Lehmkuhl, A.~Lozano-Dur{\'a}n, and J.~Rabault.
\newblock Flow control in wings and discovery of novel approaches via deep
  reinforcement learning.
\newblock \emph{Fluids}, 2022.
\newblock \doi{10.3390/fluids7020062}.

\bibitem[Vinuesa et~al.(2024)Vinuesa, Rabault, Azizpour, and
  Guastoni]{vinuesa2024stateobservation}
R.~Vinuesa, J.~Rabault, H.~Azizpour, and L.~Guastoni.
\newblock Influence of the state observation on deep-reinforcement-learning
  drag-reduction policies in wall-bounded flows.
\newblock In \emph{Proceedings of the 13th International Symposium on
  Turbulence and Shear Flow Phenomena (TSFP-13)}, 2024.

\bibitem[W{\"a}lchli et~al.(2024)W{\"a}lchli, Guastoni, Vinuesa, and
  Koumoutsakos]{walchli2024minimalchannel}
D.~W{\"a}lchli, L.~Guastoni, R.~Vinuesa, and P.~Koumoutsakos.
\newblock Drag reduction in a minimal channel flow with scientific multi-agent
  reinforcement learning.
\newblock \emph{Journal of Physics: Conference Series}, 2753:\penalty0 012024,
  2024.

\bibitem[Waleffe(1997)]{waleffe1997ssp}
F.~Waleffe.
\newblock On a self-sustaining process in shear flows.
\newblock \emph{Physics of Fluids}, 9\penalty0 (4):\penalty0 883--900, 1997.

\bibitem[Wallace et~al.(1972)Wallace, Eckelmann, and
  Brodkey]{wallace1972wallregion}
J.~M. Wallace, H.~Eckelmann, and R.~S. Brodkey.
\newblock The wall region in turbulent shear flow.
\newblock \emph{Journal of Fluid Mechanics}, 54\penalty0 (1):\penalty0 39--48,
  1972.

\bibitem[Zhou et~al.(2025)Zhou, Zhang, and Zhu]{zhou2025drlhighre}
Z.~Zhou, M.~Zhang, and X.~Zhu.
\newblock Reinforcement-learning-based control of turbulent channel flows at
  high {R}eynolds numbers.
\newblock \emph{Journal of Fluid Mechanics}, 1006:\penalty0 A12, 2025.
\newblock \doi{10.1017/jfm.2025.27}.

\end{thebibliography}

\end{document}